\documentclass[reprint,amsmath,amssymb,aps,prm,superscriptaddress,showkeys]{revtex4-2}
\usepackage{graphicx}               
\usepackage[colorlinks=true,linkcolor=blue,citecolor=black,urlcolor=black]{hyperref} 
\usepackage{xcolor}
\usepackage{amsmath,amssymb}

\makeatletter
\newcommand*{\Rmnum}[1]{\expandafter\@slowromancap\romannumeral #1@}
\newcommand*{\rmnum}[1]{\romannumeral#1\relax}
\makeatother
\newcommand{\br}{\mathbf{r}}

\begin{document}	

\title{Curvature-Controlled Band Alignment Transitions in 1D van der Waals
Heterostructures}

\author{Shu Zhao}
\affiliation{School of Materials Science and Engineering, Zhejiang
  University, Hangzhou 310027, China}
\affiliation{Key Laboratory of 3D Micro/Nano Fabrication and
  Characterization of Zhejiang Province, School of Engineering,
  Westlake University, Hangzhou 310024, China}
\affiliation{Institute of Advanced Technology, Westlake Institute for Advanced Study, Hangzhou 310024, China}

\author{Chunxia Yang}
\affiliation{Key Laboratory of 3D Micro/Nano
  Fabrication and Characterization of Zhejiang Province, School of
  Engineering, Westlake University, Hangzhou 310024, China}
\affiliation{Institute of Advanced Technology, Westlake Institute
  for Advanced Study, Hangzhou 310024, China}

\author{Ziye Zhu}
\affiliation{Key Laboratory of 3D Micro/Nano
  Fabrication and Characterization of Zhejiang Province, School of
  Engineering, Westlake University, Hangzhou 310024, China}
\affiliation{Institute of Advanced Technology, Westlake Institute
  for Advanced Study, Hangzhou 310024, China}

\author{Xiaoping Yao}
\affiliation{Key Laboratory of 3D Micro/Nano
  Fabrication and Characterization of Zhejiang Province, School of
  Engineering, Westlake University, Hangzhou 310024, China}
\affiliation{Institute of Advanced Technology, Westlake Institute
  for Advanced Study, Hangzhou 310024, China}

\author{Wenbin Li}
\email{liwenbin@westlake.edu.cn}
\affiliation{Key Laboratory of 3D Micro/Nano Fabrication and
  Characterization of Zhejiang Province, School of Engineering,
  Westlake University, Hangzhou 310024, China}
\affiliation{Institute of Advanced Technology, Westlake Institute
  for Advanced Study, Hangzhou 310024, China}

\date{\today}


  
\begin{abstract}  
One-dimensional (1D) van der Waals (vdW) heterostructures, formed
between coaxial nanotubes of transition metal dichalcogenides (TMDCs),
have emerged as a new area of endeavor in nanoscience. A key to
designing and engineering the properties of such 1D vdW
heterostructures lies on understanding the band alignment of coaxial
nanotubes in the heterostructures. However, how curvature, tube
diameters, and intertube coupling affect the band-edge levels and band
alignment of TMDC nanotubes in 1D vdW heterostructures remains
unknown. Here, through comprehensive first-principles calculations and
analyses, we establish a complete framework of band alignment in 1D
vdW heterostructures of TMDC nanotubes.  We reveal that, as the
diameter of a TMDC nanotube decreases, the combined
effects of curvature-induced flexoelectricity and intrinsic
circumferential tensile strain cause a rapid and continuous lowering
of the conduction band minimum (CBM), whereas the valence band maximum
(VBM) exhibits an initial lowering before rising, which originates
from a change in the orbital character of the VBM. The transition in
the orbital character of VBM also leads to direct-to-indirect bandgap
transition in small-diameter armchair and chiral nanotubes, as well as
photoluminescence quenching in zigzag nanotubes. As individual TMDC
nanotubes form coaxial 1D vdW heterostructures, the effect of
intertube coupling via flexovoltage effect can result in a transition
of intertube band alignment from Type~\Rmnum{2} to Type~\Rmnum{1} in
multiple heterostructural systems, including large-diameter
MoSe$_2$@WS$_2$, MoTe$_2$@MoSe$_2$, and MoTe$_2$@WS$_2$
heterostructures. These results lay down a foundation for the rational
design of 1D vdW heterostructures.
\end{abstract}

\maketitle

\section*{Introduction} 

Band alignment and band energy diagram are core concepts of
semiconductor physics that explain a wide range of phenomena
underlying applications such as transistors and quantum well
lasers~\cite{Kroemer2001,Sze2008}. The key ingredients for determining
the band alignment at a semiconductor interface include the energy
levels of the valence band maximum (VBM) and conduction band minimum
(CBM) of each semiconductor, as well as possible interfacial coupling
effects. In addition to successes in studying bulk semiconductor
interfaces, band alignment has demonstrated its predictive power in
the study of low-dimensional semiconductor heterostructures such as
stacked two-dimensional (2D) materials~\cite{Geim2013}, where
interlayer excitons were observed in van der Waals (vdW)
heterostructures of 2D transition metal dichalcogenides (TMDC) with a
staggered (Type~\Rmnum{2}) band alignment~\cite{Hong2014, Rivera2015,
  Rivera2018}. Recently, one-dimensional (1D) vdW heterostructures of
coaxial nanotubes have emerged as an intriguing new class of
nanomaterials~\cite{Xiang2020, Xiang2021, Cambre2021, Feng2021,
  Zheng2021, Guo2021}. Single-walled, single-crystal MoS$_2$ nanotubes
with diameters as small as 3.9 nm have been grown on single-walled
carbon nanotubes (SWCNTs) and boron-nitride (BN) nanotubes using
chemical vapor deposition~\cite{Xiang2020}. This provides a fresh new
ground for discovering unique phenomena and functionalities in these
novel heterostructures. Understanding the band alignment between
coaxial nanotubes in 1D vdW heterostructures will be a critical step
in this endeavour.

A prerequisite to determining the band alignment in 1D vdW
heterostructures lies on resolving the diameter-dependent evolution of
the band-edge levels of individual nanotubes. In this respect,
although there have been abundant studies of the diameter-dependent
electronic properties of SWCNTs~\cite{Hamada1992, Saito1992,
  Mintmire1992, White1993, Kane1997, Gulseren2002, Shan2005,
  Dresselhaus1998, Charlier2007}, similar investigations of TMDC
nanotubes are surprisingly limited~\cite{Seifert2000, Seifert2000a,
  Seifert2000b, Wu2007, Milosevic2007, Zibouche2012, Milivojevic2020,
  Ghosh2020, Wang2021, Mikkelsen2021, Hisama2021}. In particular, a
systematic understanding of the curvature- and diameter-dependent
band-edge energy levels and characters of TMDC nanotubes is still
lacking.

\begin{figure*}[t!]
  \centering \includegraphics[width=0.72\textwidth]{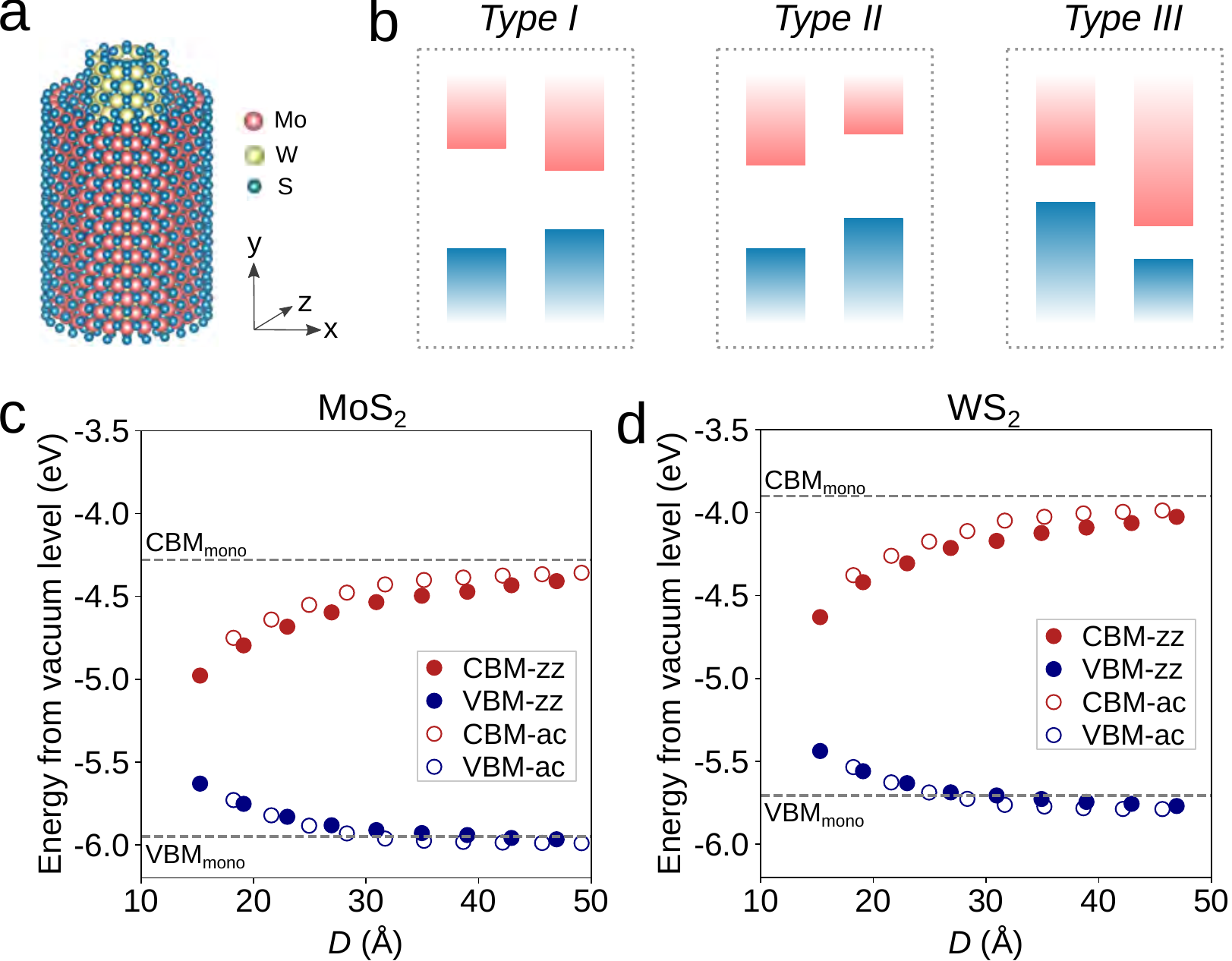}
  \caption{One-dimensional (1D) van der Waals (vdW)
    heterostructures and diameter-dependent band-edge levels of
    MoS$_2$ and WS$_2$ nanotubes. (\textbf{a}) Atomistic model of a
    WS$_2$@MoS$_2$ 1D vdW heterostructure, which corresponds to a
    smaller-diameter WS$_2$ nanotube nested in a larger-diameter
    MoS$_2$ nanotube. (\textbf{b}) Three types of band alignment in
    semiconductor heterostructures: Type~\Rmnum{1} (straddling),
    Type~\Rmnum{2} (staggered), and Type~\Rmnum{3}
    (broken-gap). (\textbf{c},\textbf{d}) Band-edge energy levels of
    MoS$_2$ nanotubes (\textbf{c}) and WS$_2$ nanotubes (\textbf{d})
    as a function of tube diameter ($D$), as calculated from density
    functional theory (DFT). The energies are measured from the vacuum
    level. The conduction band minimum (CBM) and valence band maximum
    (VBM) levels of zigzag (zz) and armchair (ac) nanotubes are
    represented by CBM-zz, VBM-zz, CBM-ac and VBM-ac,
    respectively. The gray dashed lines illustrate the DFT-calculated
    CBM and VBM levels of monolayer systems.}
  \label{fig:fig1}
\end{figure*}

The above considerations motivate us to establish a complete framework
of band alignment in 1D vdW heterostructures of TMDC nanotubes. We
first carry out a comprehensive investigation of the
diameter-dependent VBM and CBM levels of semiconducting Mo- and
W-dichalcogenide nanotubes, which include MoS$_2$, MoSe$_2$, MoTe$_2$,
WS$_2$, and WSe$_2$ nanotubes. The remaining member in this material
class, namely WTe$_2$, has a semi-metallic ground
state~\cite{Wilson1969}. We employ density functional theory (DFT)
calculations to determine the absolute energy levels of the VBM and
CBM of TMDC nanotubes in reference to vacuum level. This common vacuum
energy reference allows us to directly compare the band-edge levels of
nanotubes of varying diameters. Our DFT calculations focus on zigzag
and armchair nanotubes with diameters that ranges from 1.5~nm to 5~nm,
the upper end of which exceeds those of experimentally synthesized
single-walled MoS$_2$ nanotubes grown on SWCNTs and BN nanotubes
(3.9~nm)~\cite{Xiang2020}. Additional computational details are
described in the Method section.

We find that the band-edge levels of TMDC nanotubes exhibit a highly
non-trivial dependence on the tube diameters. When the tube diameter
is above 50~\AA, curvature-induced flexoelectricity and its associated
electrostatic potential effect is found to be the main reason that
affects the band-edge levels of individual TMDC nanotubes. Below a
diameter value of $\sim$50~\AA, the circumferential tensile strain
inherent in TMDC nanotubes starts to play an increasingly important
and eventually dominant role in the band-edge evolution, causing a
rapid lowering of the CBM of TMDC nanotubes as well as a transition
from downward to upward shifting in VBM levels as tube diameter
decreases. The transition in diameter-dependent evolution of VBM
levels is caused by a strain-induced change in the VBM character from
in-plane to out-of-plane orbitals, and this transition has profound
implications on the electronic and optical properties of TMDC
nanotubes, including direct-to-indirect bandgap transition in armchair
and chiral nanotubes, as well as photoluminescence quenching in zigzag
nanotubes. Quantum confinement is found to play a negligible role on
the band-edge evolution.

After obtaining an in-depth understanding of the band-edge evolution
of individual TMDC nanotubes, we have studied the band alignment of
coaxial TMDC nanotubes in 1D vdW heterostructures, taking into account
of the effect of intertube electronic coupling. We find that the
flexovoltage generated by the outer nanotube on the inner nanotube in
a 1D vdW heterostructure plays a critical role in the band alignment
of coaxial nanotubes. Combining the results on individual nanotubes
and intertube coupling effects, we establish a complete framework of
band alignment in all 1D vdW heterostructures formed between TMDC
nanotubes, such as the WS$_2$@MoS$_2$ 1D vdW heterostructures
illustrated in Figure~\ref{fig:fig1}a. We find that as the tube
diameters of 1D vdW heterostructures decrease, a transition from
Type~\Rmnum{2} to Type~\Rmnum{1} band alignment
(Figure~\ref{fig:fig1}b) can occur in multiple heterostructural
systems. In particular, Type~\Rmnum{2} to Type~\Rmnum{1}
band-alignment transitions can occur in large-diameter
MoSe$_2$@WS$_2$, MoTe$_2$@MoSe$_2$, and MoTe$_2$@WS$_2$
heterostructures, making these heterostructural systems attractive for
nanoscale optoelectronic applications.

  
\section*{Results and Discussion}

We first present our results and analyses of the diameter-dependent
band-edge level evolution in TMDC nanotubes. The DFT-calculated CBM
and VBM levels of MoS$_2$ and WS$_2$ nanotubes are presented in
Figure~\ref{fig:fig1}c,d. The results reveal that the CBM levels of
MoS$_2$ nanotubes exhibit a continuous lowering with the decrease of
tube diameter ($D$). In contrast, the VBM levels exhibit a more
complex change. In the limit of infinitely large diameter, the tube
curvature approaches zero and therefore the VBM level shall be equal
to that of the corresponding monolayer. However, the VBM levels
computed for MoS$_2$ and WS$_2$ nanotubes with diameters around
50~\AA\ are lower than those of the corresponding monolayers,
indicating an initial lowering of the VBM levels as the tube diameters
decrease.  The VBM of MoS$_2$ nanotubes already starts to show a
slowly upward shift when the diameter is below $\sim$50~\AA, whereas
in WS$_2$ nanotubes, the upward shift becomes noticeable at a diameter
of around 40~\AA. When the tube diameter is below 30~\AA, the CBM
level lowers rapidly with a further reduction of the tube diameter,
whereas the VBM level exhibits the opposite trend, raising in a fast
pace with decreasing tube diameter.  These trends are the same for
both zigzag and armchair tubes, indicating weak chirality dependence
of band-edge level evolution.

\begin{figure*}[t!]
  \centering \includegraphics[width=1.0\textwidth]{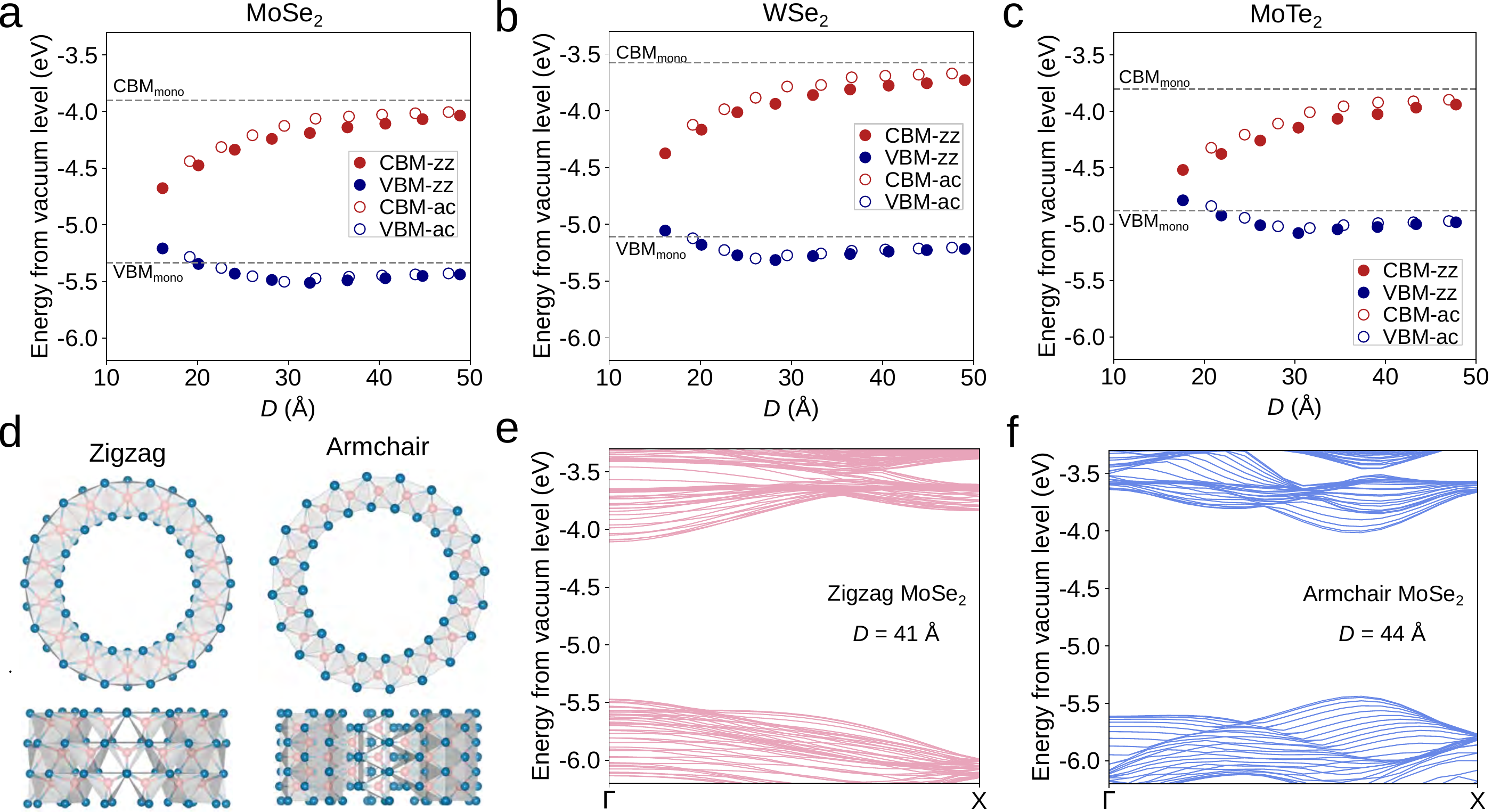}
  \caption{(\textbf{a}-\textbf{c}) Evolution of the VBM and CBM of
    MoSe$_2$, WSe$_2$, and MoTe$_2$ nanotubes as a function of the
    tube diameters. (\textbf{d}) Atomistic representations of zigzag
    and armchair nanotubes of transition metal dichalcogenides
    (TMDCs), viewed along the tube axis (top panels) and from the side
    (bottom panels). The pink and cyan spheres represent
    transition-metal and chalcogen atoms,
    respectively. (\textbf{e},\textbf{f}) Electronic band structures
    of zigzag (\textbf{e}) and armchair (\textbf{f}) nanotubes with a
    diameter $\sim$40~\AA. The electron wave vectors are along the
    direction of tube axis, and only half of the 1D Brillouin zone is
    shown due to symmetry.}
  \label{fig:fig2}
\end{figure*}

The general trend of the variation in the VBM and CBM levels of
MoSe$_2$, WSe$_2$, and MoTe$_2$ nanotubes are similar to MoS$_2$ and
WS$_2$, but the non-monotonic change of their VBM is even more
evident, as shown in Figure~\ref{fig:fig2}a--c. Notably, the VBM of
the three Se- and Te-based TMDC nanotubes do not rise until the tube
diameter is below $\sim$30 \AA. The transition diameters are
$\sim$30~\AA, 28~\AA, and 30~\AA\ for MoSe$_2$, WSe$_2$, and MoTe$_2$
nanotubes, respectively. The maximum lowering of the VBM in reference
to the corresponding monolayers, reached at the ``transition
diameters'', are on the order of 0.2~eV.

Since VBM and CBM correspond to the edges of the valence and
conduction bands, respectively, it is informative to inspect the
electronic band structures of TMDC nanotubes. The atomistic structures
of zigzag and armchair nanotubes are illustrated in
Figure~\ref{fig:fig2}d. The calculated electronic band structures of a zigzag
MoSe$_2$ nanotube with diameter $D=41$~\AA\ and an armchair MoSe$_2$
nanotube of $D=44$~\AA\ are shown in Figure~\ref{fig:fig2}e,f. At
these diameter values, both zigzag and armchair MoSe$_2$ nanotubes are
direct bandgap semiconductors. In the zigzag MoSe$_2$ nanotubes, both the
VBM and CBM reside at the $\Gamma$ point, whereas in the armchair
nanotubes, the band edges are located at around $2/3$ of the path from
$\Gamma$ to $X$, where $X$ is the 1D Brillouin zone
boundary. DFT-calculated results indicate that, as the tube diameters
are reduced further, zigzag nanotubes will remain their direct-bandgap
nature. By contrast, a direct-to-indirect bandgap transition will
occur in armchair nanotubes, due to the relative upward shift of the
electronic states near $\Gamma$ (Figure~S1). The
direct-to-indirect bandgap transition is observed in all armchair
nanotubes of Mo-
and W-dichalcogenides (MoS$_2$, MoSe$_2$, MoTe$_2$, WS$_2$,
and WSe$_2$). Such a transition has previously been observed in
MoS$_2$ and MoSe$_2$ nanotubes, where the diameters corresponding to
the transitions were determined to be 52~\AA\ and 33~\AA,
respectively, in close agreement with our
results~\cite{Wu2018,Hisama2021}.

To investigate the origin of the diameter-dependent band-edge levels
of TMDC nanotubes, we note that when a TMDC monolayer is wrapped into
a nanotube, three distinct factors could affect the band-edge levels:
(\rmnum{1}) The electron orbitals are spatially confined along the
circumferential direction due to the small tube diameter, leading to
possible quantum confinement effect. (\rmnum{2}) The radial curvature
of the nanotube can generate flexoelectricity and flexoelectric
potential~\cite{Shan2005, Artyukhov2020, Springolo2021} inside the
nanotube, which could shift the energy levels of the electronic
states. (\rmnum{3}) The bending of a 2D sheet into a nanotube and the
subsequent structural relaxation alter the crystal symmetry and
generate internal tensile and bending strains in the nanotube, which
change the electron orbital hybridization and thus the band-edge
levels.

\begin{figure*}[t!]
  \centering \includegraphics[width=1.0\textwidth]{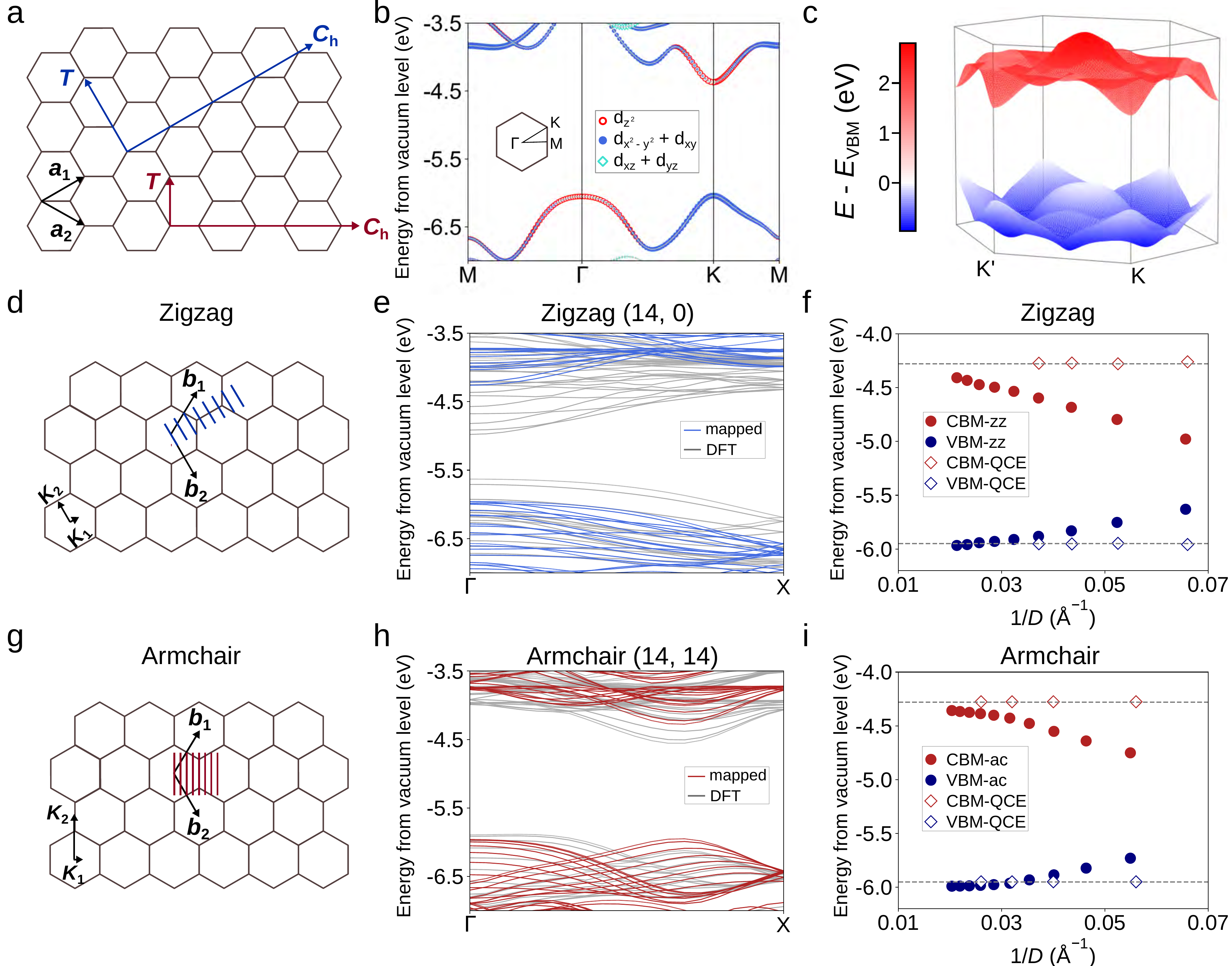}
  \caption{Effect of circumferential quantum confinement on the
    band-edge levels of MoS$_2$ nanotubes. (\textbf{a}) Top view of
    the direct lattice of a monolayer MoS$_2$. The 2D primitive
    lattice vectors are indicated by $\mathbf{a}_1$ and
    $\mathbf{a}_2$. The chiral vector $\mathbf{C}_h$ and the
    corresponding translation vector $\mathbf{T}$ of zigzag nanotubes
    (shown in blue) and armchair nanotubes (shown in red) are
    illustrated. (\textbf{b}) 2D band structure of monolayer
    MoS$_2$. The relative contributions of the Mo $d_{z^2}$,
    $d_{x^2-y^2} + d_{xy}$, and $d_{xz}+d_{yz}$ orbitals to the
    electronic states are illustrated. (\textbf{c}) Full-zone 3D band
    structure of monolayer MoS$_2$. (\textbf{d}) 2D-to-1D band
    structure mapping of zigzag nanotubes. The reciprocal lattice
    vectors of monolayer MoS$_2$ are indicated by $\mathbf{b}_1$ and
    $\mathbf{b}_2$. $\mathbf{K}_1$ and $\mathbf{K}_2$ are the
    reciprocal lattice vectors of $\mathbf{C}_h$ and $\mathbf{T}$. The
    blue line segments correspond to the unique crystal momentums that
    satisfy the circumferential periodic boundary conditions of a
    zigzag nanotube. The separation between two segments equals
    $|\mathbf{K}_1|$, and the length of each line segment is
    $|\mathbf{K}_2|$. For a zigzag nanotube with chiral index $(n,0)$,
    the number of line segments is $2n$. The hypothetical 1D band
    structure of a nanotube can be obtained by cutting the band
    structure of a 2D monolayer along the blue line segments and mapped
    to 1D along the direction of $\mathbf{K}_2$.  (\textbf{e}) Mapped
    and DFT-calculated band structures of a zigzag (14,0) nanotube,
    shown as blue and gray curves, respectively. (\textbf{f}) The CBM
    and VBM levels of mapped 1D band structures are shown as open diamond
    markers. The DFT calculated results are represented by filled
    circles. (\textbf{g},\textbf{h},\textbf{i}) Similar to
    (\textbf{d},\textbf{e},\textbf{f}), but for armchair nanotubes
    with chiral indices $(n,n)$. The number of line segments in
    (\textbf{g}) is also $2n$.}
  \label{fig:fig3}
\end{figure*}

\textbf{Quantum Confinement Effect (QCE).} We first consider the
effect of circumferential QCE in TMDC nanotubes by using the
Brillouin-zone folding scheme that was originally developed for
studying QCE in SWCNTs~\cite{Dresselhaus1998}. QCE has been shown to
have a strong influence on the band-edge levels of
SWCNTs~\cite{Hamada1992, Saito1992, Mintmire1992, White1993, Kane1997,
  Shan2005}. In Mo- and W-dichalcogenide 2D semiconductors, QCE in the
layer-normal direction is also the key factor that drives
indirect-to-direct bandgap transition from bulk to
monolayer~\cite{Mak2010,Splendiani2010}.

From structural perspective, a TMDC nanotube can be thought of as
derived from a monolayer TMDC, which is rolled up and seamed to form a
tubular structure with a finite radial curvature. The direction of
rolling and the circumference of the resultant tube are determined by
the ``chiral vector" $\mathbf{C}_h = n \mathbf{a}_1 + m$$
\mathbf{a}_2$, where $\mathbf{a}_1$ and $\mathbf{a}_2$ are the 2D
primitive lattice vectors of the corresponding TMDC
monolayers~\cite{Dresselhaus1998}, as illustrated in
Figure~\ref{fig:fig3}a. The two integer indices $n$ and $m$ form a
pair $(n,m)$ that specifies the chirality of the tube, with $(n, n)$
corresponding to armchair tubes, $(n\neq0, m=0)$ and its symmetric
pair $(n=0, m\neq0)$ zigzag tubes, and other cases of $(n \neq m, m
\neq 0)$ being chiral tubes. The translational vector
$\mathbf{T}$, which is parallel to the tube axis and normal to
$\mathbf{C}_h$, reflects the 1D translational symmetry of the nanotube.

Without loss of generality, here we use MoS$_2$ as a representative to
discuss the effect of QCE on the band-edge level evolution of TMDC
nanotubes. In the zone-folding scheme~\cite{Dresselhaus1998}, we first
calculate the 2D electronic band structure of monolayer MoS$_2$, as
shown in Figure~\ref{fig:fig3}b,c. Bloch electron wavevectors that
satisfy the periodic boundary conditions imposed by the formation of
nanotubes are then determined, which leads to the discretization of
monolayer Brillouin zone into 1D line segments. From the discretized
Brillouin zone, the 2D electronic states are mapped to 1D, resulting
in the 1D bandstructures of nanotubes within the zone-folding scheme.

In more technical terms, a nanotube can be unrolled into a
hypothetical curvatureless nanoribbon. In this unrolled ribbon, the
Bloch wavefunction $\Psi_{\mathbf{k}}(\mathbf{r})$ must find itself
after traveling a distance corresponding to the chiral vector
$\mathbf{C}_h$, that is, $\Psi_{\mathbf{k}}(\mathbf{r} + \mathbf{C}_h)
= \Psi_{\mathbf{k}}(\mathbf{r})$, where $\mathbf{k}$ is a 2D crystal
momentum. In combination with the Bloch's theorem, the boundary
conditions require that the phase of electron wavefuctions gained due
to a translation of $\mathbf{C}_h$, that is, $\exp(i\mathbf{k}\cdot
\mathbf{C}_h)$, be unity. This leads to $\mathbf{k}\cdot \mathbf{C}_h
= 2\pi \mu$, where $\mu$ is an integer.

The reciprocal vectors $\mathbf{K}_1$ and $\mathbf{K}_2$ that
correspond to $\mathbf{C}_h$ and $\mathbf{T}$ can be defined by
following standard orthogonality relations between real and reciprocal
lattices: $\mathbf{K}_1 \cdot \mathbf{C}_h = 2\pi$, $\mathbf{K}_1
\cdot \mathbf{T} = 0$, $\mathbf{K}_2 \cdot \mathbf{C}_h = 0$, and
$\mathbf{K}_2 \cdot \mathbf{T} = 2\pi$. Using these relations, the
crystal momentum $\mathbf{k}$ that satisfies $\mathbf{k}\cdot
\mathbf{C}_h = 2\pi \mu$ can be written in terms of $\mathbf{K}_1$ and
$\mathbf{K}_2$ as:
\begin{equation}
  \mathbf{k} = k \frac{\mathbf{K}_2}{|\mathbf{K}_2|} + \mu
  \mathbf{K}_1,
  \label{eq:eq1}
\end{equation}
where $k$ is a continuous real number. Due to the translational
symmetry of nanotubes along $\mathbf{T}$, the value of $k$ in
eq.~\ref{eq:eq1} can be restricted between $-|\mathbf{K_2}|/2$ and
$|\mathbf{K_2}|/2$, that is, between $-\pi/T$ and $\pi/T$, where $T$
equals $|\mathbf{T}|$.  Furthermore, denoted by $N$ the number of TMDC
formula units within the area enclosed by $\mathbf{C}_h$ and
$\mathbf{T}$, it can be proven that $N\mathbf{K}_1$ corresponds to a
reciprocal lattice vector of the 2D
monolayer~\cite{Dresselhaus1998}. As wavevectors differ by a
reciprocal lattice vector are equivalent, the choices of $\mu$ can be
restricted to between $0$ and $N-1$. These considerations lead to the
following formula for mapping the hypothetical 1D band structure
$E_{\textrm{1D}}(k)$ of a nanotube from the 2D band structure
$E_{\textrm{2D}}(\mathbf{k})$ of monolayer ~\cite{Dresselhaus1998}:
\begin{equation}
\begin{gathered}
  E_{\text{1D}}^\mu(k) = E_{2D}(k\frac{\mathbf{K}_2}{|\mathbf{K}_2|} + \mu
  \mathbf{K}_1), \\
  (-\frac{\pi}{T} \leq k < \frac{\pi}{T},
  \text{and} \ \mu = 0, ..., N-1).
\end{gathered}
\label{eq:band_map}
\end{equation}
For each 1D crystal momentum $k$, there are $N$ branches. Each branch
represents a slice of the corresponding 2D band structure.

We can further express $\mathbf{K}_1$ and $\mathbf{K}_2$ in the above
equation (eq.~\ref{eq:band_map}) in terms of the 2D reciprocal lattice
vectors of monolayer, namely $\mathbf{b}_1$ and $\mathbf{b}_2$.  It
can be proven~\cite{Dresselhaus1998} that for a zigzag $(n,0)$
nanotube, $N = 2n$, $\mathbf{K}_1 = (2\mathbf{b}_1 + \mathbf{b}_2)/N$,
and $\mathbf{K}_2 = -\mathbf{b}_2/2$, as illustrated in
Figure~\ref{fig:fig3}d. On the other hand, for an armchair nanotube
with the chiral index $(n,n)$, $N$ is still equal to $2n$, but
$\mathbf{K}_1 = (\mathbf{b}_1 + \mathbf{b}_2)/N$, and $\mathbf{K}_2 =
(\mathbf{b}_1 - \mathbf{b}_2)/2$, as shown in Figure~\ref{fig:fig3}g.

Eq.~\ref{eq:band_map} now represents a complete formula for mapping
the 1D band structure of a nanotube from that of a corresponding 2D
monolayer. We shall emphasize that the band structure of nanotubes
thus obtained only consider the effect of quantum confinement of
electronic states along the circumferences of nanotubes. Following
this scheme, we have calculated the mapped band structures of a zigzag
(14,0) and an armchair (14,14) nanotubes, whose diameters are
15~\AA\ and 25~\AA, respectively. The results are shown in
Figure~\ref{fig:fig3}e and Figure~\ref{fig:fig3}h, respectively. In
these figures, DFT-calculated band structures are plotted together
with the mapped ones for comparison. It can be seen that, although the
mapped and DFT-calculated band structures share similar features in
terms of band dispersion, the band-edge levels and bandgaps exhibit a
significant difference. Furthermore, if we plot the band-edge levels
obtained from the zone-folding scheme as a function of tube diameter,
as shown in Figure~\ref{fig:fig3}f and Figure~\ref{fig:fig3}i, they
exhibit little variation with respect to inverse diameter $1/D$. This
indicates that QCE plays a negligible role in the band-edge level
evolution of both zigzag and armchair nanotubes.

The above result can be understood on the basis of that, even for a
zigzag (14,0) with a diameter as small as 15~\AA, there are already 28
line segments ($N = 2n$) cutting through the 2D band Brillouin
zone. Since monolayer MoS$_2$ and other 2D TMDCs are semiconductors
with a bandgap around 1--3~eV as well as relatively large electron and hole
effective masses~\cite{Ramasubramaniam2012, Peelaers2012}, the 1D line
segments can sample 2D electron states whose energies are very close
to the VBM and CBM of a 2D monolayer. In fact, for armchair $(n,n)$
nanotubes, it is can be seen from Figure~\ref{fig:fig3}g that one of
the line segments is guaranteed to pass through both the $K$ and $K'$
points of the 2D hexagonal Brillouin zone, where the CBM and VBM of
monolayer TMDC reside. This results in exactly the same band-edge
levels of the mapped 1D and the original 2D systems for armchair
nanotubes. For zigzag nanotubes of very small diameters (15~\AA), a
very small amount of raising of CBM and lowering of the VBM is 
observable in Figure~\ref{fig:fig3}f, but the trend is opposite to the
DFT-calculated results in this diameter regime.

Although the 2D-to-1D band-structure mapping leads to the conclusion
that circumferential QCE has little influence on the band-edge levels
of nanotubes, it does provide important insight into the electronic
structure of TMDC nanotubes. As a key example, the band-structure
mapping explains why in zigzag nanotubes, both the VBM and CBM are
always located at the $\Gamma$ point of the 1D Brillouin zone. In
fact, it can be seen from Figure~\ref{fig:fig3}d that, for zigzag
nanotubes, within the zone-folding scheme, the line segments for
band-structure mapping pass through both the $\Gamma$ and $K$ valleys
of 2D Brillouin zone at the centers of the line segments. Hence, when
plotting out the 1D band structure, the highest $\Gamma$-valley and
$K$-valley derived valence-band energies, as well as the lowest
$K$-valley derived conduction-band energies, will all be mapped to the
$\Gamma$ points of the 1D Brillouin zones of zigzag nanotubes
(Figure~\ref{fig:fig3}e). As a result, zigzag nanotubes are expected
to have both VBM and CBM located at $\Gamma$. If we consider effects
beyond QCE as perturbations to the mapped 1D band structures, zigzag
nanotubes are expected to be direct-bandgap semiconductors regardless
of tube diameters, in consistent with DFT results.

Since the valence-band energies of TMDC monolayers at $\Gamma$ and $K$
are quite close ($E_{\text{VB}}^{K} - E_{\text{VB}}^{\Gamma}$ around
0.1~eV for MoS$_2$ monolayer~\cite{Jin2013} and 0.38~eV for MoSe$_2$
monolayer~\cite{Zhang2014}), if effects beyond QCE are considered, the
VBM of zigzag nanotubes may either derive from the valence-band
electronic states of 2D monolayer at or near the $K$ valley, which
have mainly in-plane Mo $d_{x^2-y^2} + d_{xy}$ and S $p_x+p_y$
character, or derive from monolayer states at $\Gamma$, which mainly
have Mo $d_{z^2}$ and S $p_z$ contributions (see also
Figure~S2)~\cite{Cao2012, Zhu2011}.  The CBM of zigzag
nanotubes are expected to derive from the conduction-band states in
the $K$-valley electronic states of monolayer, which mainly have Mo
$d_{z^2}$ character with a small S $p_x+p_y$ contribution.

On the other hand, in armchair nanotubes, we can see from
Figure~\ref{fig:fig3}g that, after band-structure mapping, $K$-derived
electronic states are separated from $\Gamma$-derived states in the
new 1D Brillouin zone. The $K$-derived states will be located at 2/3
of the $\Gamma$-$X$ path, while the $\Gamma$-derived states remain at
the $\Gamma$ of the 1D Brillouin zone of a nanotube. Hence, if only
QCE effect is considered, armchair nanotubes of TMDCs are
direct-bandgap semiconductors with both VBM and CBM located at 2/3 of
the $\Gamma$-$X$ path. However, effects beyond QCE may lead to the
raising of the energy levels of $\Gamma$-derived electronic states
with respect to $K$-derived states. In this case, the VBM of nanotubes
will shift to $\Gamma$ (or the vicinity of $\Gamma$), and the system
may become an indirect-bandgap semiconductor. This is exactly what
happens to armchair nanotubes with small tube diameter
(Figure~\ref{fig:fig3}h).

Now that QCE has been excluded as a major contribution to
diameter-dependent band-edge level in TMDC nanotubes, our study will
focus on how the band-edges, which derive from the electronic states
of monolayer at $\Gamma$ or $K/K'$ valleys, respond to the effects of
flexoelectricity and strain in the nanotubes.

\begin{figure*}[t!]
  \centering
  \includegraphics[width=1.0\textwidth]{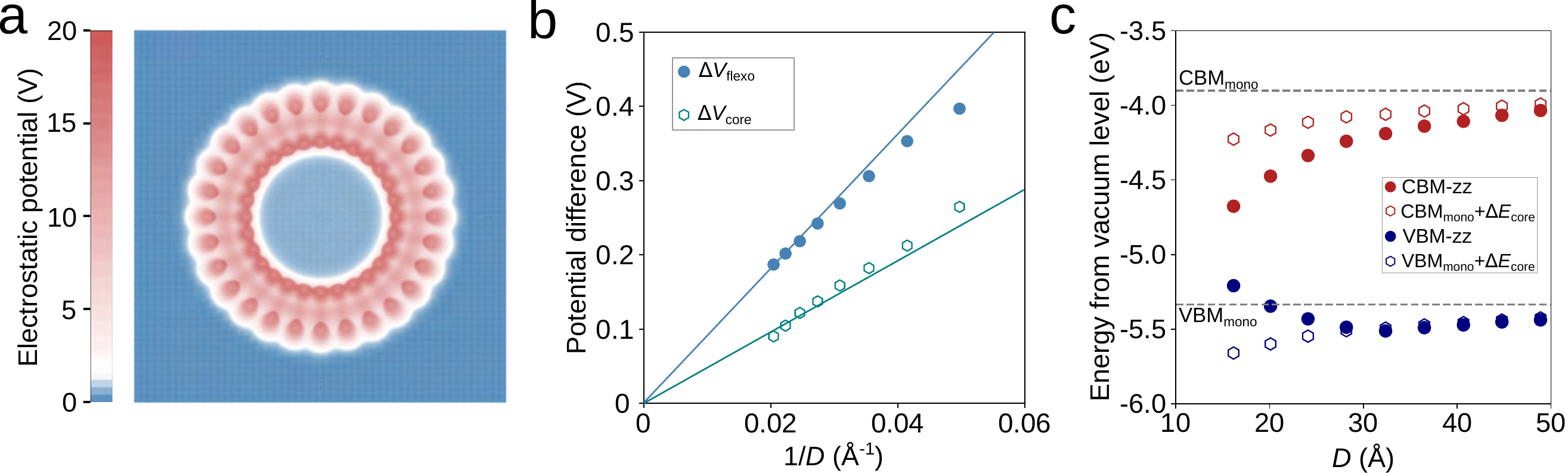}
  \caption{Effect of curvature-induced flexoelectricity on the
    band-edge levels of MoSe$_2$ nanotubes. (\textbf{a})
    DFT-calculated electrostatic potential profile in a MoSe$_2$
    nanotube. The potential is averaged along the axial direction of
    the tube and projected to a 2D cross-section plane. The
    electrostatic potential in the vacuum region inside the tube
    ($V_{\textrm{inner}}$) is higher than the vacuum potential outside
    the tube ($V_{\textrm{outer}}$). The voltage difference, namely
    the flexovoltage, can be computed as $\Delta V_{\textrm{flexo}} =
    V_{\textrm{inner}} - V_{\textrm{outer}}$. In this particular
    MoSe$_2$ zigzag (14,0) nanotube with a diameter of $\sim$16~\AA,
    $\Delta V_{\textrm{flexo}}$ is around 0.43~V. (\textbf{b})
    Calculated flexovoltages (filled blue dots) and average local
    electrostatic potential changes at the Mo site ($\Delta
    V_{\textrm{core}}$, open green circles) as a function of 
    inverse diameter $1/D$.  The lines are linear extrapolations of
    each set of data in the large-diameter limit. (\textbf{c}) Effect
    of flexoelectricity-induced electrostatic potential energy shift
    on the band-edge level evolution of MoSe$_2$ zigzag nanotubes. The
    filled circles are the VBM and CBM levels of MoSe$_2$ nanotubes
    from actual DFT calculations. The open circles represent the model
    band-edge levels of MoSe$_2$ nanotubes obtained by adding the
    VBM/CBM energy levels of MoSe$_2$ monolayer (VBM$_{\text{mono}}$
    and CBM$_{\text{mono}}$) with the average electrostatic potential
    energy shift ($\Delta E_{\textrm{core}} = -|e|\Delta
    V_{\textrm{core}}$, where $e$ is the elementary charge) at the
    corresponding tube diameter.}
  \label{fig:fig4}
\end{figure*}

\textbf{Curvature-Induced Flexoelectric and Electrostatic Potential
  Effect.}  When a sheet of 2D material is rolled to form a nanotube,
the resulting curvature leads to the redistribution of charge inside
and outside the nanotube, generating radial electric
polarizations~\cite{Dumitrica2002,Kvashnin2015, Shi2018,
  Artyukhov2020, Kumar2021, Codony2021, Bennett2021,
  Springolo2021}. This phenomenon has essentially the same origin as
flexoelectricity in bulk crystals, where a strain gradient breaks
local inversion symmetry and generates macroscopic
polarization~\cite{Wang2019}. In fact, in reference to a flat TMDC
monolayer, the outer chalcogen layer of a TMDC nanotube is stretched,
whereas the inner layer is compressed. As a result, in the
neighborhood of the surface of a TMDC nanotube, an effective
transverse strain gradient of $1/R$ exists in the layer normal
direction~\cite{Springolo2021}, where $R$ is the radius of
the nanotube. The resulting radial flexoelectric polarization generates
an electrostatic potential difference between the inner and outer
sides of the nanotube, as illustrated for a MoSe$_2$ zigzag $(14,0)$
nanotube in Figure~\ref{fig:fig4}a. The voltage difference is named
flexovoltage and denoted by $\Delta V_{\textrm{flexo}}$, which can be
computed as $\Delta V_{\textrm{flexo}} = V_{\textrm{inner}} -
V_{\textrm{outer}}$. Here $V_{\textrm{inner}}$ and
$V_{\textrm{outer}}$ represent the electrostatic potentials of the
hollow region inside the tube and the vacuum region far away from the tube
surface, respectively. At the leading order, $\Delta
V_{\textrm{flexo}}$ is proportional to inverse diameter
$1/D$~\cite{Dumitrica2002,Springolo2021}.

The flexoelectric polarization and the corresponding redistribution of
charge also change the average electrostatic potential experienced by the
electronic states inside a nanotube. This effect will lead to a shift
in the absolute energy levels of the electronic states with respect to
the vacuum. For a given atomic arrangement and electron density
distribution, the local electrostatic potential inside a nanotube is
given by $V_{\mathrm{elstat}}(\br) = V_{\mathrm{ion}}(\br) -
\frac{|e|}{4\pi \epsilon_0} \int \frac{n(\br')}{|\br - \br'|} d\br'$,
where $V_{\mathrm{ion}}(\br)$ is the ionic potential from all atoms,
and the second term is the Hartree potential from electrons with
density distribution $n(\br)$.  $e$ is the elementary charge, and
$\epsilon_0$ is the vacuum permittivity. The
$V_{\mathrm{elstat}}(\br)$ of a MoSe$_2$ zigzag $(14,0)$ nanotube is
visualized in Figure~\ref{fig:fig4}a.

The average electrostatic potential experienced by an electrostatic
state inside a nanotube, whose normalized electron wavefunction given
by $\psi(\br)$, can be calculated as $\bar{V}_{\mathrm{elstat}} = \int
V_{\mathrm{elstat}}(\br)|\psi(\br)|^2 d\br$. Since the band-edge
states of TMDC monolayers and nanotubes predominantly derive from the
$d$ orbitals of the transition-metal atoms~\cite{Zhu2011, Cao2012,
  Liu2013}, the change in $\bar{V}_{\mathrm{elstat}}$ of a band-edge
state before and after forming a nanotube, denoted by $\Delta
\bar{V}_{\mathrm{elstat}}$, can be approximated by the average local
electrostatic potential shift near the transition-metal site ($\Delta
V_{\textrm{core}}$). $\Delta V_{\textrm{core}}$ is calculated within a
spherical region centered at the lattice site, with a radius of
$\sim$1.2~\AA, which is roughly half of the nearest-neighbor distance
between Mo and Se. The corresponding electrostatic potential energy
shift, $\Delta E_{\text{core}} = -|e|\Delta V_{\textrm{core}}$, is
similar in magnitude to the energy level shift of the core electrons
in the transition-metal atoms, as the core electrons are tightly bound
to the nuclei and do not participate in chemical bonding. We have
also calculated the average local electrostatic potential shifts of
all atoms (including both Mo and Se) in the tube, and found that the
difference between the two approaches is small (with a relative
difference of less than 10~\%), indicating the validity of our
approach. In other words, $\Delta V_{\textrm{core}}$ calculated at the
transition-metal site can be well approximate the
$\bar{V}_{\mathrm{elstat}}$ of band-edge states in TMDC nanotubes.

Figure~\ref{fig:fig4}b shows the calculated values of $\Delta
V_{\textrm{core}}$, together with the values of $\Delta
V_{\textrm{flexo}}$, for MoSe$_2$ zigzag nanotubes as a function of
inverse diameter $1/D$. The calculated $\Delta V_{\textrm{flexo}}$ and
$\Delta V_{\textrm{core}}$ of MoS$_2$, WS$_2$, WSe$_2$, and MoTe$_2$
nanotubes are of similar values, as shown in Figure~S3. For
TMDC tubes with diameters between 50~\AA\ to 15~\AA, $\Delta
V_{\textrm{flexo}}$ are in the range from 0.2~V to 0.4~V, which are
three to four times larger than those of SWCNTs at similar
diameters~\cite{Springolo2021}, indicating quite strong flexoelectric
effects in TMDC nanotubes.

Our result in Figure~\ref{fig:fig4}b also indicates
that, in the large-diameter limit, $\Delta V_{\textrm{core}}$ is
around one half of $\Delta V_{\textrm{flexo}}$ at the corresponding
diameter. This is understandable given that flexoelectricity-induced
electrostatic potential change decreases from the inner to outer side
of a tube, eventually becoming zero far outside the tube.

With the calculated electrostatic potential change, we can determine
the contribution of curvature-induced flexoelectricity on the
band-edge levels of TMDC nanotubes. To this end we compute the
hypothetical VBM and CBM levels of TMDC nanotubes by adding
the calculated electrostatic potential energy shift $\Delta E_{\text{core}} =
-|e|\Delta V_{\textrm{core}}$ to the CBM and VBM energies of
the monolayer, and then compare these model-derived band-edge levels to
the DFT-computed VBM and CBM levels of nanotubes. The results for
MoSe$_2$ zigzag nanotubes are shown in Figure~\ref{fig:fig4}c.

It can be concluded from Figure~\ref{fig:fig4}c that, when tube
diameters are around or larger than 50~\AA, the shifts in the energy
levels of the CBM and VBM of MoSe$_2$ nanotubes with respect to that
of the monolayer can be explained by the curvature-induced flexoelectric
effect. At smaller diameters ($D$ below 30~\AA\ for VBM and below
50~\AA\ for the CBM of MoSe$_2$ nanotubes), however, additional
contributions beyond flexoelectricity must be taken into account to
explain the evolution of band-edge levels.  In particular, the
transition from downward to upward shifting in the VBM levels, when
$D$ is below 30~\AA, cannot be rationalized within the framework of
flexoelectric effect.

\textbf{Circumferential Tensile Strain Effect.} Given that
flexoelectricity-induced electrostatic effect alone cannot explain the
evolution of VBM and CBM in small-diameter tubes, we next investigate
the effect of strain on the band-edge levels. As mentioned earlier,
when a TMDC monolayer is rolled into a nanotube, the inner layer of
chalcogen atoms are compressed while the outer-layer atoms are
stretched. This causes a distortion to the trigonal prismatic
coordination between the transition-metal and chalcogen atoms. The
distortion becomes increasingly severe as the tube diameter decreases,
resulting in a rapid increase of bending energy~\cite{Seifert2002}.
The bending energy in the system, however, can be partially reduced if
the tube diameter is slightly enlarged, albeit at the cost of adding
circumferential tensile strain energy into the system. That is,
enlarging the circumference of a nanotube through bond stretching
leads to a decrease in the curvature and thus the bending energy, at
the cost of adding tensile strain energy along the circumferential
direction of the tube.

\begin{figure*}[t!]
	\centering
	\includegraphics[width=1.0\textwidth]{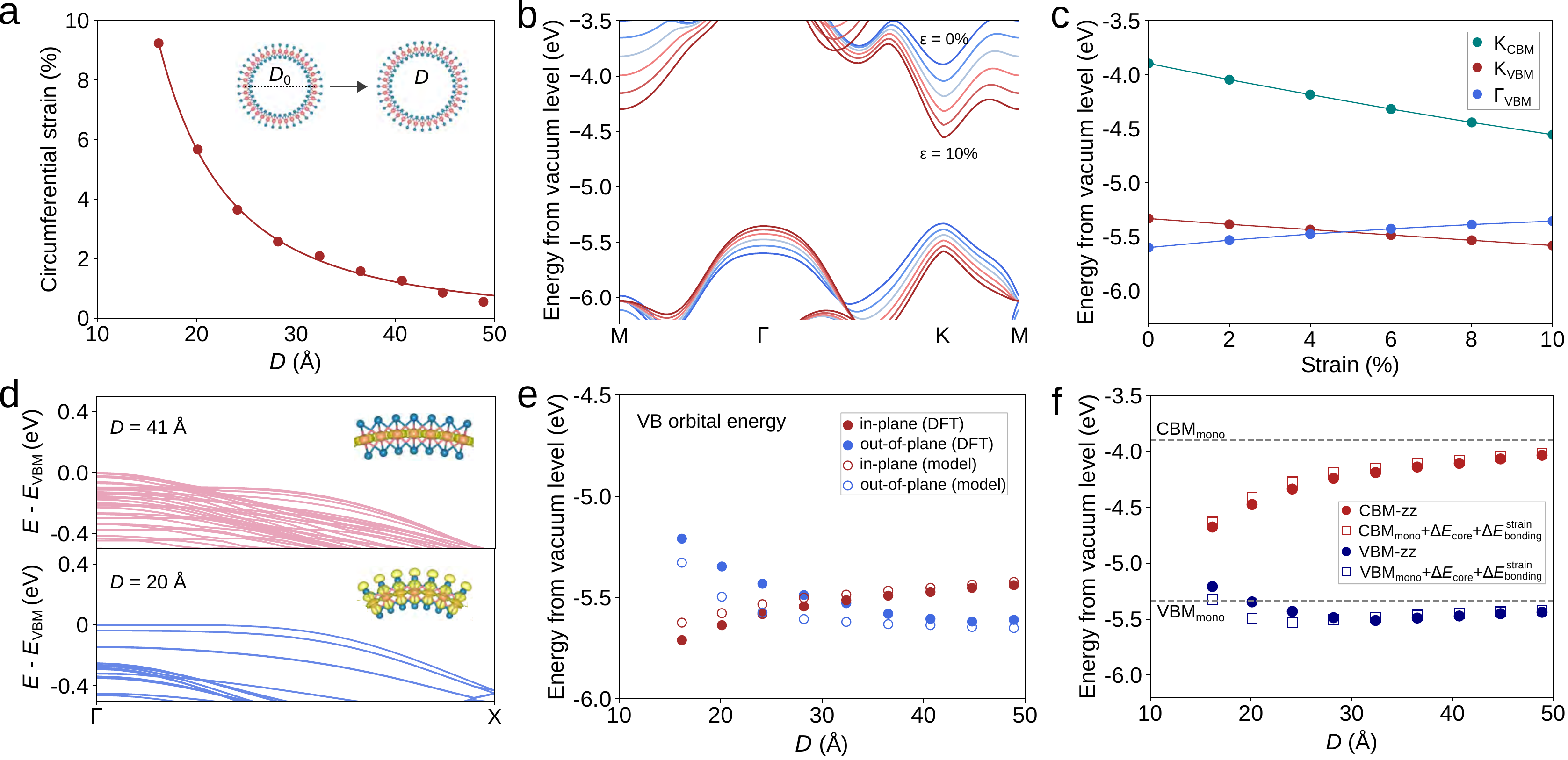}
	\caption{Effect of circumferential tensile strain on the
          band-edge levels of MoSe$_2$ nanotubes. (\textbf{a})
          DFT-computed circumferential tensile strain as a function of
          tube diameter $D$. The reference for the strain calculation
          at each tube diameter is the circumference of a
          corresponding unrelaxed tube rolled from a monolayer. The
          circumference of a tube is measured from the
          transition-metal atoms. As illustrated, after structural
          relaxation $D > D_0$, where $D_0$ is the original tube
          diameter. For easier visualization, the enlargement of tube
          diameter in the atomistic model of the inset is slightly
          exaggerated. (\textbf{b}) Evolution of the electronic band
          structure of a MoSe$_2$ monolayer as a function of uniaxial
          tensile strain. The strain varies from zero to 10\% at a
          step of 2\%. The electron orbital energies are aligned with
          respect to vacuum level. (\textbf{c}) Plot of the band-edge
          energy levels of MoSe$_2$ monolayer as a function of
          uniaxial tensile strain. The green, red, and blue lines
          represent the data for CBM at $K$, VBM at $K$, and VBM at
          $\Gamma$, respectively. (\textbf{d}) Change in the VBM
          orbital character as the diameter of zigzag nanotube is
          decreased. The top panel shows the electronic band structure
          near the valence band edge of a zigzag MoSe$_2$ tube with a
          diameter of 41~\AA. The inset on the top right illustrates
          the electron density isosurface of the VBM state at
          $\Gamma$. The bottom panel shows the corresponding results
          for a tube with a smaller diameter of 20~\AA. The VBM of the
          larger-diameter tube has an in-plane character (Mo $d_{x^2 -
            y^2}$ + $d_{xy}$ derived), whereas in the smaller-diameter
          tube, the VBM has an out-of-plane character (Mo $d_{z^2}$
          and Se $p_z$ derived). (\textbf{e}) Diameter dependence of
          the valence-band (VB) orbital energy, for the topmost
          in-plane and out-of-plane orbitals. The filled circles are
          from DFT calculations, whereas the open circles are from our
          model that takes into account both the flexoelectric and
          circumferential tensile strain effects.  (\textbf{f})
          Comparison of the DFT-computed CBM and VBM energies of
          MoSe$_2$ zigzag nanotubes (filled circles) with those from
          our model (open squares). $\Delta E_{\text{core}}$ as
          defined earlier represents the averaged
          electrostatic-potential energy shifts near the Mo lattice
          site, and $\Delta E_{\text{bonding}}^{\text{strain}}$
          represents the bonding part of the contribution to
          circumferential tensile strain-induced band-edge shifts (see
          main text).}
\label{fig:fig5}
\end{figure*}

The competition between bending energy (which favors a larger tube
diameter) and circumferential tensile strain energy (which favors a
smaller deviation from the original tube diameter) results in an
optimized diameter that is slightly larger than the original value of
$D_0 = |\mathbf{C}_h|/\pi$, where $\mathbf{C}_h$ is
the chiral vector defined in a pristine monolayer before
rolling. In Figure~\ref{fig:fig5}a we illustrate the enlargement of
the diameter of a MoSe$_2$ nanotube after structural relaxation in
DFT, and present the circumferential tensile strains in MoSe$_2$
nanotubes as a function of relaxed tube diameters $D$. The
circumferential tensile strain $\varepsilon$ is defined as the
relative change in the circumference of a tube before and after
structural relaxation: $\varepsilon = (\pi D - \pi D_0)/\pi D_0$, that
is, $\varepsilon = D/D_0 -1$. Both $D$ and $D_0$ are measured from the
transition-metal atoms. A rapid increase in $\varepsilon$ when $D$ is
below 50~\AA\ can be observed. The strain versus diameter curve in
Figure~\ref{fig:fig5}a can be well fitted by $\varepsilon = \alpha
(d/D)^2 + \beta (d/D)^{3}$, where $d = 6.5$~\AA\ is the interlayer
distance in bulk MoSe$_2$~\cite{Bronsema1986}, and $\alpha = 0.39$,
$\beta = 3.0$ are best-fit numerical coefficients.

The above scaling of circumferential tensile strain $\varepsilon$ with
respect to tube diameter $D$ can be rationalized within the framework
of continuum mechanics. For a rectangular sheet of solid with
thickness $d$, after rolling into a tube, its bending energy per unit
volume $E_{\textrm{bending}}$ scales with $1/D$ as
$E_{\textrm{bending}} = \frac{1}{6} Y (d/D)^2$ at the leading order,
where $Y$ is the in-plane Young's modulus~\cite{LandauBook}. For
better connection to the continuum limit, we take the thickness $d$ of
TMDC monolayers as the interlayer distance of the corresponding
bulk. On the other hand, the circumferential tensile strain energy per
unit volume $E_{\textrm{tensile}}$, at the leading order, is given by
$E_{\textrm{tensile}} = \frac{1}{2} Y \varepsilon^2$. The total strain
energy per unit volume $E_{\textrm{total}}$ is then the sum of
$E_{\textrm{bending}}$ and $E_{\textrm{tensile}}$: $E_{\textrm{total}}
= \frac{1}{6}Y (d/D)^2 + \frac{1}{2} Y \varepsilon^2$. Since
$\varepsilon = D/D_0 -1$, $\varepsilon = 0$ if $D = D_0$. However,
$\varepsilon = 0$ is not the lowest-energy state: $E_{\textrm{total}}$
can be reduced if $D$ is slightly larger than $D_0$. Minimizing
$E_{\textrm{total}}$ with respect to $D$, we obtain $\varepsilon =
\frac{d^2}{3}\frac{D_0}{D^3} \approx \frac{1}{3}\left( \frac{d}{D}
\right)^2$. The coefficient of the second-order term in $d/D$ of the
$\varepsilon$ versus $D$ scaling from this simple analysis is $1/3$,
which is in close agreement with our earlier DFT numerical result of
$\alpha = 0.39$.

The above analysis establishes that both circumferential tensile
strain and bending strain exist in TMDC nanotubes. Both types of
strain are expected to affect the band-edge levels through
strain-induced changes in the electron-orbital interaction and
hybridization. This ``chemical bonding contribution'' adds to the
electrostatic contribution from the flexoelectric effect discussed
earlier. While the flexoelectric potential shift from
curvature-induced flexoelectricity in general causes a lowering of the
energy levels of both VBM and CBM with respect to
vacuum energy, strain-induced band-edge energy shift can be either
positive or negative, depending on the nature of orbital interaction
involved in a specific band-edge state.

We first investigate the effect of circumferential tensile strain on
the band-edge levels of TMDC nanotubes (the effect of bending strain
will be discussed later). This effect can be studied by first looking
at the evolution of the band-edge levels of TMDC monolayers under uniaxial
tensile strain, and then employing the zone-folding framework
developed earlier to deduce the corresponding circumferential
strain-induced changes in the band-edge levels of nanotubes.

We again use MoSe$_2$ monolayers and nanotubes as model systems to
illustrate the effect of circumferential tensile strain on the
evolution of band-edge states. The strain-dependent electronic
structures of TMDC monolayers have been investigated
before~\cite{Feng2012,Yun2012,Johari2012,Wiktor2016}, but its
connection to the band-edge evolution of TMDC nanotubes has not been
carefully studied. Figure~\ref{fig:fig5}b,c show that, as uniaxial
tensile strain is imposed along the zigzag direction of monolayer,
both the absolute energy levels of VBM and CBM at the $K$ point of 2D
Brillouin zone exhibit a downward shift. In contrast, the VBM at
$\Gamma$ point exhibits an upward shift.

The different strain-dependent behavior among the three band-edge
states of TMDC monolayers can be rationalized within the tight-binding
picture of solid-state electronic structure, in terms of the
competition between on-site orbital energy and inter-site orbital
hopping energy. The on-site orbital energy is affected by the local
electrostatic potential, while the inter-site orbital hopping is
sensitive to the distance between neighboring atoms and influences the
splitting between bonding and anti-bonding orbitals. The band-edge
states of TMDC monolayers are predominantly derived from the $d$
orbitals of transition-metal atoms~\cite{Liu2013,Fang2015}. As tensile
strain causes the increase of the distance between orbitals on
neighboring atomic sites, the hopping integrals are in general
reduced, leading to a smaller bonding-antibonding splitting. This
contributes to a raise in the absolute energy of a bonding orbital and
a lowering in the energy of an antibonding orbital.

On the other hand, tensile strain also changes the local electrostatic
environment of electron orbitals on each atoms, which affects their
on-site energies. This effect is akin to the crystal-field effect in
chemistry. As in-plane tensile strain is imposed on a TMDC monolayer,
the increased interatomic distance causes an overall decrease in the
electron density surrounding the core region of each atom (since the
contribution to electron density from other atoms is reduced). This
leads to an increase in the local electrostatic potential at atomic
sites and hence a decrease in the electrostatic potential energies of
electron orbitals residing on the atoms. The strain-induced local
electrostatic potential change is indeed found in DFT calculations:
Figure~S4 shows that in a strained MoSe$_2$ monolayer, the
electrostatic potential energy of an electron residing at the Mo site
decreases almost linearly with uniaxial tensile strain at a rate of
$\sim$30~meV per one percent of tensile strain.

Therefore, whether a band-edge state in a monolayer raises or lowers
in energy after imposing a tensile strain depends on its
bonding/antibonding nature, as well as the competition between bonding
effect (inter-site hopping integral) and electrostatic effect (on-site
energy). An antibonding orbital will decrease in energy as the two
effects add up together, whereas a bonding orbital may either exhibit
upward or downward shifting in the absolute energy, depending on which
of the two effects is more dominant. In many semiconductors (but not
all), the VBM is a bonding orbital while the CBM is an anti-bonding
orbital. Hence, the absolute band-edge shifts under strain is
typically larger in CBM than in VBM, as illustrated schematically in
Figure~S5.

In a TMDC monolayer, the CBM at $K$ is an antibonding orbital that
mainly derives from the transition-metal $d_{z^2}$ orbitals and, to a 
lesser extent, the chalcogen $p_x + p_y$
orbitals~\cite{Cao2012}. Hence, when a tensile strain is applied, this
antibonding orbital (CBM at $K$) rapidly lowers in energy. In
contrast, the top valence-band (VB) states at $K$ and $\Gamma$ are
bonding orbitals. The former (VBM at $K$) has an in-plane orbital
character that mainly derives from the transition-metal $d_{x^2-y^2} +
d_{xy}$ and chalcogen $p_x + p_y$ orbitals, whereas the latter (VBM at
$\Gamma$) has an out-of-plane transition-metal $d_{z^2}$ and chalcogen
$p_z$ character~\cite{Cao2012, Zhu2011}. DFT results in
Figure~\ref{fig:fig5}b show that, under uniaxial tensile strain, the
energy of the VBM at $K$ exhibits a downward shift (negative
deformation potential~\cite{Wiktor2016}), whereas the VBM at $\Gamma$
exhibits an upward shift (positive deformation
potential). Nevertheless, the absolute values of VB energy shifts at
$K$ and $\Gamma$ are both smaller than that of CBM at $K$, in
consistent with the competition between bonding/antibonding effect and
electrostatic effect in the strain-induced variation of bonding
orbitals.

When the uniaxial tensile strain imposed on a MoSe$_2$ monolayer is
beyond a critical value ($\sim$4.9\%), the energy of VBM at $\Gamma$
becomes higher than that of VBM at $K$, resulting in a
direct-to-indirect bandgap transition in MoSe$_2$ monolayer
(Figure~\ref{fig:fig5}c). Similar strain-induced band-edge evolution
also occurs in other TMDCs monolayers, and the critical strain of
direct-to-indirect bandgap transition increases in the order of
MoS$_2$, WS$_2$, MoSe$_2$, WSe$_2$, and MoTe$_2$~\cite{Johari2012,
  Wiktor2016}. This order is mainly due to the larger $K$-$\Gamma$
energy separations in the valence bands of selenide monolayers than
those in the sulfide counterparts~\cite{Zhu2011,Jin2013, Zhang2014},
as the slopes of strain-dependent VBM energies are quite similar among
different TMDC monolayers~\cite{Wiktor2016}.  Besides, although the
results presented here are for TMDC monolayers uniaxially stretched
along the zigzag direction, previous studies have shown that, due to
the in-plane isotropy of TMDC monolayers, uniaxial strain along other
directions (such as the armchair direction) leads to almost identical
results of absolute band-edge shifts in TMDC
monolayers~\cite{Johari2012, Wiktor2016}.

On the basis of tensile-strain induced band-edge evolution in TMDC
monolayers, we can invoke the zone-folding scheme of monolayer-to-tube
electronic-structure mapping as discussed earlier, and expect that the
circumferential tensile strain in TMDC nanotubes would induce
alterations in the corresponding mapped band-edge state. With respect
to armchair nanotubes, since both the VBM and CBM at $K$ in monolayer
are mapped to $2/3$ of the $\Gamma$-$X$ path in the 1D Brillouin zone
of nanotube ($k = {2\pi}/{3T}$), whereas the VB state at $\Gamma$ in
monolayer is mapped to the $\Gamma$ of nanotube, it is expected that
in armchair TMDC nanotubes, as diameter decreases, circumferential
tensile strain will cause downward shifts of both CBM and VBM at $k =
{2\pi}/{3T}$, but an upward shift of the VBM at $\Gamma$. When the
diameter of a tube is sufficiently small, circumferential tensile
strain will eventually cause the energy level of VBM at $\Gamma$ to
become higher than that of the VBM at $k = {2\pi}/{3T}$, rendering
small-diameter armchair nanotubes indirect bandgap
semiconductors. This direct-to-indirect bandgap transition of armchair
nanotubes was indeed observed in
earlier~\cite{Seifert2000,Wu2018,Hisama2021} and our DFT calculations
(Figure~S1). Accompanying this direct-to-indirect bandgap
transition is a change in the VBM from $K$ valley-derived to $\Gamma$
valley-derived.

On the other hand, when it comes to zigzag TMDC nanotubes, since the
monolayer VBM and CBM states at (or near) $K$, as well as the
monolayer VBM state at $\Gamma$, are all mapped to the $\Gamma$ point
of 1D Brillouin zone of the corresponding nanotube, zigzag nanotubes
are expected to remain direct-bandgap semiconductors regardless of the
tube diameter. Nevertheless, the VBM of large-diameter zigzag
nanotubes are expected to derive from the monolayer VBM at $K$ and
have similar orbital characters, whereas the VBM of small-diameter
zigzag nanotubes are expected to derive from monolayer VBM at
$\Gamma$, since in a uniaxially strained monolayer, the energy level
of VBM at $\Gamma$ eventually becomes higher than that of VBM at $K$.
Hence, a change in the orbital character of VBM will also occur in
small-diameter zigzag nanotubes. Before the occurrence of this change
in band-edge orbital character, the absolute energy level of VBM in
zigzag nanotubes is expected to exhibit a downward shift as the
tube diameter decreases, whereas after the transition in orbital
character, decreasing tube diameter will raise the VBM energy. Thus,
despite the chirality difference, the diameter dependence of the
band-edge levels of armchair and zigzag nanotubes exhibit the same
trend, as indeed observed in our DFT calculations
(Figure~\ref{fig:fig1}c,d and Figure~\ref{fig:fig2}a--c).

The change in the orbital character of VBM in MoSe$_2$ zigzag
nanotubes from large-diameter to small-diameter ones is shown in
Figure~\ref{fig:fig5}d. We can see from the figure that, when the
diameter of a MoSe$_2$ zigzag nanotube is $\sim$41 \AA, the VBM has an
in-plane orbital character with mostly Mo $d_{x^2-y^2} + d_{xy}$
contribution. In contrast, in a smaller-diameter tube with $D = 20$
\AA, the VBM has an out-of-plane character with Mo $d_{z^2}$ and Se
$p_z$ character. This is in consistent with the zone-folding of zigzag
nanotubes from a strained monolayer. In Figure~\ref{fig:fig5}e, we
further track the energies of topmost in-plane and out-of-plane VB
orbitals as a function of tube diameter, and find that the transition
of VBM character from in-plane $d_{x^2-y^2} + d_{xy}$ to out-of-plane
$d_{z^2}$ orbitals occurs at a diameter of $\sim$30~\AA. This is in
line with the fact that, at this diameter, the VBM energy levels of
MoSe$_2$ nanotubes start to increase with the decrease of tube
diameters.

The above results allow us to quantitatively model the
diameter-dependent band-edge shifts of TMDC nanotubes by taking into
account both flexoelectric effect and circumferential tensile
strain effect.  To this end, we first revisit Figure~\ref{fig:fig4}b,
where we calculate the shift of average electrostatic potential near
the Mo site in MoSe$_2$ nanotubes with respect to that of MoSe$_2$
monolayer, and we show in Figure~\ref{fig:fig4}c that the
electrostatic potential shifts cause a lowering of both VBM and CBM in
MoSe$_2$ nanotubes. These results, however, were calculated from
relaxed TMDC nanotubes, which possess circumferential tensile strain
after structural optimization. However, as concluded from our earlier
discussion of strained monolayer, in addition to curvature-induced
flexoelectric potential effect, the bond-distance change induced by
circumferential tensile strain can also cause an electrostatic
potential energy shift in TMDC nanotubes (Figure~S4). Hence,
if we want to account for the effects of flexoelectric potential and
circumferential tensile strain together, directly adding the total
band-edge energy shifts caused by circumferential tensile strain (as
modeled using uniaxially tensile strained monolayers) on top of the
total electrostatic potential energy shifts of nanotubes in
Figure~\ref{fig:fig4}c may cause a double counting of electrostatic
contribution. Instead, the correct approach would be adding only the
bonding part of the contribution $\Delta
E_{\text{bonding}}^{\text{strain}}$ from circumferential tensile
strain. $\Delta E_{\text{bonding}}^{\text{strain}}$ can be obtained by
subtracting the tensile strain-induced electrostatic-potential energy
shift $\Delta E_{\text{core}}^{\text{strain}}$ from the total 
energy-level shift of VBM or CBM in monolayers, as illustrated in
Figure~S5. The results of $\Delta
E_{\text{bonding}}^{\text{strain}}$ for uniaxially strained MoSe$_2$
monolayers are presented in Figure~S6.

To compare the flexoelectricity-induced and circumferential tensile
strain-induced electrostatic potential energy shifts, we carried out
the following computational experiment: when relaxing the structure of
a MoSe$_2$ nanotube rolled from a 2D monolayer, the positions of
transition-metal atoms are fixed in space, while chalcogen atoms are
allowed to relax in DFT. This fixes the diameter of the tube during
structural relaxation. The nanotube after the restricted relaxation
will thus not possess circumferential tensile strain. We then compute
the electrostatic potential energy shifts near the Mo site in such
restrictively relaxed nanotubes (zero circumferential tensile strain),
using the same methodology as in fully relaxed nanotubes. The result,
shown in Figure~S7, indicates that for tube diameter above
50~\AA, the flexoelectric effect dominates the electrostatic potential
energy shifts, as circumferential tensile strain is small in this
diameter regime (Figure~\ref{fig:fig5}a). However, for tube diameter
below 40~\AA, the circumferential tensile strain-induced electrostatic
potential energy shift is comparable or even larger than the
flexoelectric contribution.

Having obtained the bonding part of the circumferential tensile
strain-induced band-edge shifting $\Delta
E_{\text{bonding}}^{\text{strain}}$, we then model the VBM energy
levels of nanotubes as $\text{VBM}_{\text{tube}} =
\text{VBM}_{\text{mono}} + \Delta E_{\text{core}} + \Delta
E_{\text{bonding}}^{\text{strain}}$, where $\text{VBM}_{\text{tube}}$
and $\text{VBM}_{\text{mono}}$ represent the VBM energies of nanotube
and monolayer, respectively. The model for CBM is analogous. We then
calculate and plot the modeled VBM and CBM of MoSe$_2$ zigzag
nanotubes and compare the results to actual DFT-computed values, as
shown in Figure~\ref{fig:fig5}f (see also Figure~\ref{fig:fig5}e for a more
specific comparison between model and DFT for topmost in-plane and
out-of-plane VB orbitals). The results demonstrate that for tube
diameter above 30~\AA, combining flexoelectric effect and
circumferential tensile strain effect can well explain the evolution
of the energy levels of both VBM and CBM in MoSe$_2$ nanotubes. In
particular, the rapid upward shifts in VBM energies at small tube
diameter are correctly captured.

\textbf{Bending Strain Effect.} While our model combining
flexoelectric and circumferential tensile strain effects can well
explain the evolution of band-edge levels in TMDC nanotubes, it can be
seen from Figure~\ref{fig:fig5}f that the model prediction becomes
less quantitatively accurate as tube diameter becomes smaller,
especially when the tube diameter is below the critical point of
transition in VBM orbital character. This is because, as tube diameter
decreases, bending strain will start to play a more important role in
the band-edge evolution, whose effect has not been included in our
model.  Indeed, when the tube diameter is below the transition
diameter, both the VBM and CBM of TMDC nanotubes have mainly
transition-metal $d_{z^2}$ characters. At the leading order, curvature
will reduce the hopping integral $t$ between $d_{z^2}$ orbitals on
adjacent sites as $t = t_0 - \gamma \left( \frac{a}{R} \right)^2$,
where $\gamma$ is a coefficient, $a$ is the distance between two
neighboring sites, and $R$ is the tube radius~\cite{Ando2000,
  delValle2011}. A decrease in $R$ will contribute to a decrease in
the hopping integral $t$. Hence, bending strain will contribute further to
decreased bonding-antibonding splitting in small-diameter tubes,
leading to additional upward shifts of VBM and downward shifts of CBM,
which is in consistent with the results in Figure~\ref{fig:fig5}f.

To sum up the preceding discussions, the diameter-dependent band-edge
evolution of TMDC nanotubes can be well understood by taking into
account the effects of flexoelectricity, circumferential tensile
strain, and bending strain.  Flexoelectric potential effect dominates
in large-diameter (above 50~\AA) nanotubes, lowering both the CBM and
VBM of TMDC nanotubes with respect to the corresponding values in a
monolayer, whereas circumferential tensile and bending strain effects
are responsible for the rapid downward and upward shifts of the CBM
and VBM in small-diameter tubes, respectively. In particular,
circumferential tensile strain is responsible for the change of the
VBM character from in-plane to out-of-plane orbitals, which causes the
non-monotonic diameter-dependent variation of the VBM energy levels.

\textbf{Photoluminescence Quenching in Small-Diameter TMDC Nanotubes}.
Our study of the diameter-dependent band-edge evolution in TMDC
nanotubes also provides insight into the optical properties of these
nanotubes. A TMDC monolayer can be considered as a tube with
an infinitely large radius of curvature. The five Mo- and
W-dichalcogenides discussed in this work are all direct-bandgap
semiconductors in the monolayer form and exhibit interband
photoluminescence~\cite{Mak2010,Splendiani2010,Kuc2011,Yun2012,Jin2013,Zhang2014}. However,
on the basis of our analysis, neither armchair nor zigzag nanotubes of
TMDC are expected to exhibit photoluminescence below a critical tube
diameter. In armchair nanotubes, the disappearance of
photoluminescence is due to a direct-to-indirect bandgap transition
below the critical diameter. In zigzag nanotubes, however, it is
caused by a change in the orbital character of VBM from mainly
in-plane transition-metal $d_{x^2 - y^2} + d_{xy}$ to out-of-plane
transition-metal $d_{z^2}$ plus chalcogen $p_z$. As the CBM orbital
remains predominantly of the transition-metal $d_{z^2}$ character
regardless of the tube diameter, the dipole matrix elements between
VBM and CBM will vanish after the transition in VBM character. This is
because, for non-zero dipole matrix elements, the initial and final
states of optical transition need to have different parity, which is
no longer the case after the transition in VBM character. The
symmetry-forbidden optical transition~\cite{Milosevic2007} thus
results in the disappearance of photoluminescence in small-diameter
zigzag nanotubes. Other chiral
nanotubes are not expected to have photoluminescence below a critical
diameter of a similar value either, as based on the 2D-to-1D zone
folding discussed earlier, all small-diameter chiral TMDC nanotubes
shall become indirect bandgap semiconductors below a critical
diameter, due to circumferential tensile strain effect, as only in the
special case of zigzag nanotubes would both the $K$ and $\Gamma$
points of the 2D Brillouin zone of a corresponding monolayer be mapped
to the same $\Gamma$ point in the 1D Brillouin zone of a nanotube (see
Figure~\ref{fig:fig3}d and \ref{fig:fig3}g).

On the basis of the above discussion of the relation between band-edge
character and photoluminescence, it is clear that the diameter value
at which the photoluminescence of TMDC nanotubes disappears is the
same as when the diameter dependence of VBM energy levels changes
sign. From Figures~\ref{fig:fig1}c,d and Figures~\ref{fig:fig2}a--c, we
determine that the transition diameters in MoS$_2$, WS$_2$, MoSe$_2$,
WSe$_2$, and MoTe$_2$ nanotubes are about 55~\AA, 50~\AA, 30~\AA,
28~\AA, and 29~\AA, respectively. Previous DFT studies have only
discussed the direct-to-indirect bandgap transition in armchair
nanotubes of TMDCs, wherein the transition diameters of MoS$_2$ and
MoSe$_2$ armchair nanotubes were separately determined to be
52~\AA\ and 33~\AA~\cite{Hisama2021, Wu2018}. Experimentally, Liu et
al. carried out optical measurements of single-walled MoS$_2$ grown on
BN nanotubes and determined that photoluminescence is still present in
MoS$_2$ nanotubes with a diameter around 60--70~\AA, with which our
DFT results are consistent~\cite{Liu2021}.  Our results indicate that
MoSe$_2$, WSe$_2$, and MoTe$_2$ nanotubes have smaller critical
diameters of photoluminescence quenching than MoS$_2$ and WS$_2$
nanotubes, making them (Se- or Te-based TMDC nanotubes) more
attractive for certain optoelectronic applications.

\textbf{Band Alignment in 1D vdW Heterostructures of TMDC Nanotubes}.
Having understood the diameter-dependent evolution of the band-edge
levels and orbital characters of individual TMDC nanotubes, we next
investigate the band alignment of coaxial TMDC nanotubes in 1D vdW
heterostructures. An example of such 1D vdW heterostructure was
introduced in Figure~\ref{fig:fig1}a, where a WS$_2$@MoS$_2$
heterostructure that consists of a single-walled WS$_2$ nanotube
nested in a larger-diameter MoS$_2$ nanotube is illustrated.

It has been known that in the case of 2D vdW heterostructures formed
between semiconducting Mo- and W-dichalcogenide monolayers, with the
only exception of the stacking between WSe$_2$ and MoTe$_2$ (which
forms Type~\Rmnum{1} band alignment), all the other nine 2D vdW
heterostructures exhibit Type~\Rmnum{2} band alignment~\cite{Kang2013,
  Liang2013,Hong2014, Rivera2015}. However, for 1D vdW
heterostructures of TMDC nanotubes, we have demonstrated that tube
diameter strongly affects the band-edge levels of individual
nanotubes. Hence, when two TMDC nanotubes of different materials types
and tube diameters are brought together, the band-edge alignment
between the nanotubes will depend on the specifics of tube
diameters. Furthermore, as a smaller-diameter TMDC nanotube is nested
inside a larger-diameter one, the flexovoltage generated by the outer
tube will cause an additional shift of the band energy levels in the
inner tube~\cite{Artyukhov2020}. The diameter-dependent band-edge
levels of individual nanotubes, in combination with the flexovoltage
effect (and possibly other intertube coupling effects, which will be
discussed later), are expected to cause a significant difference
between the band alignment of TMDC nanotubes in a 1D vdW
heterostructure versus its 2D counterpart, and possibly induce
transitions between different types of band alignment. This intriguing
possibility is investigated below.

\begin{figure*}[t!]
	\centering
	\includegraphics[width=0.82\textwidth]{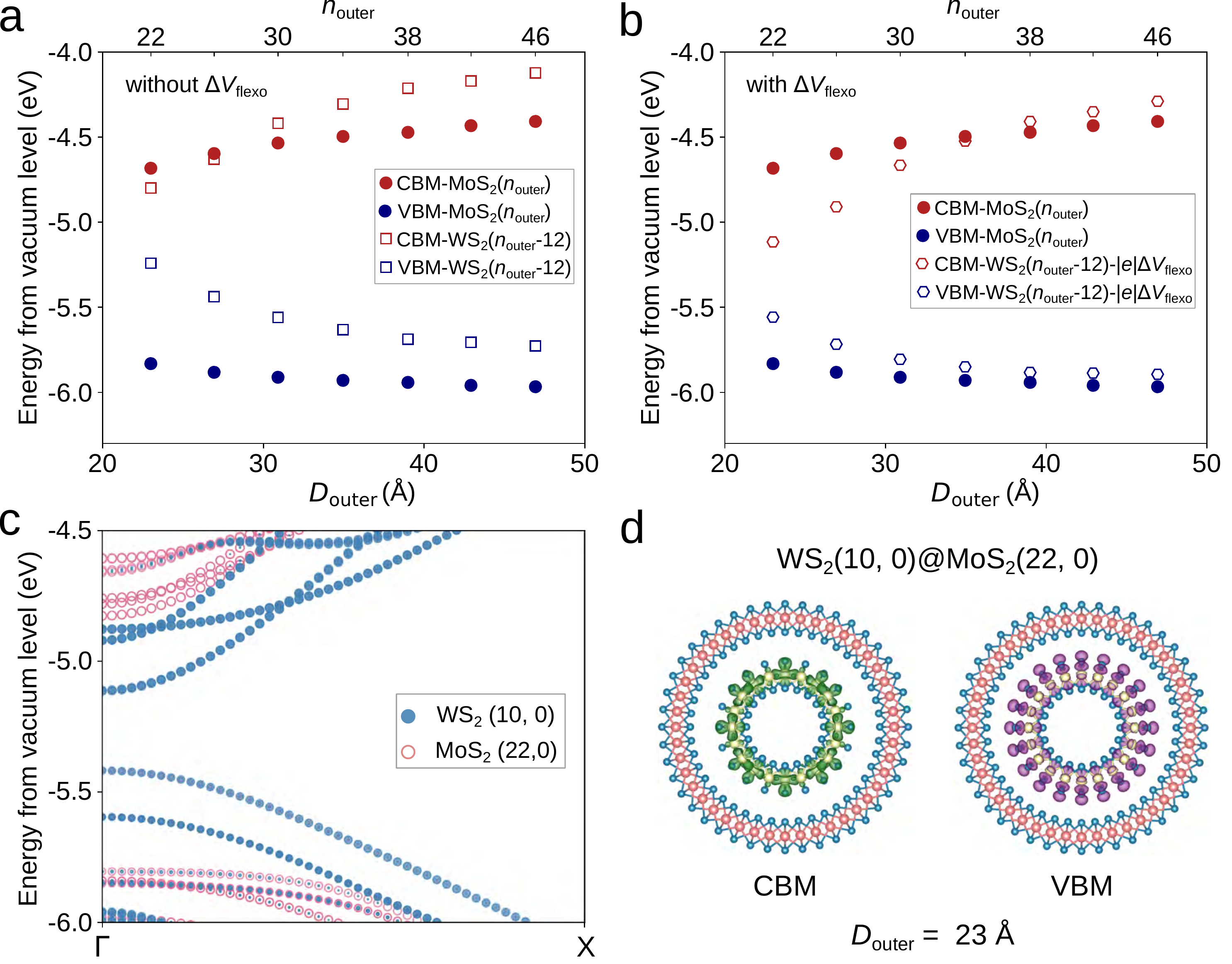}
	\caption{Band alignment in 1D vdW heterostructures consist of
          WS$_2$ nanotubes nested in MoS$_2$ nanotubes. (\textbf{a})
          CBM and VBM levels of individually separated
          MoS$_2$$(n_{\text{outer}},0)$ nanotubes and smaller-diameter
          WS$_2$$(n_{\text{outer}}-12,0)$.  $D_{\text{outer}}$ denote
          the diameters of MoS$_2$ nanotubes in the
          heterostructures. Since the tubes are separated in this
          case, the effect of flexoelectric potential $\Delta
          V_{\text{flexo}}$ on the band alignment is not
          considered. (\textbf{b}) Alignment of CBM and VBM after the
          nanotubes form coaxial WS$_2$@MoS$_2$ 1D vdW
          heterostructures. The positive flexovoltage of an outer
          MoS$_2$ nanotube leads to a downward shifting of the energy
          levels of the inner WS$_2$ nanotube by an amount of $-|e|\Delta
          V_{\text{flexo}}$. (\textbf{c}) Calculated electronic band
          structure of a WS$_2$(10,0)@MoS$_2$(22,0)
          heterostructure. Projection of electronic states to the
          inner WS$_2$ nanotube is denoted by blue filled circles and
          to the outer MoS$_2$ nanotube by red open circles, with the
          relative weight represented by the size of the
          markers. (\textbf{d}) The CBM and VBM wavefunctions
          (probability density isosurfaces) of the
          WS$_2$(10,0)@MoS$_2$(22,0) heterostructure, viewed along the
          tube-axis direction.}
\label{fig:fig6}
\end{figure*}

We first present our results on the band alignment in 1D vdW
heterostructures formed between zigzag nanotubes of MoS$_2$ and WS$_2$
(the results for heterostructures of armchair nanotubes are
found to be similar). The diameter dependence of the VBM and CBM
levels of individual MoS$_2$ and WS$_2$ nanotubes have been shown in
Figure~\ref{fig:fig1}c,d, and the flexovoltages $\Delta
V_{\text{flexo}}$ of TMDC nanotubes are  shown in
Figure~S3.

In a coaxial 1D vdW heterostructure, the diameter
of the inner nanotube is, by definition, smaller than the outer
one. As such 1D vdW heterostructures are expected to form through
templated growth of an outer tube on an inner one~\cite{Xiang2020}, or
else via elemental substitution~\cite{Zhang2017} in double-walled
nanotubes, we consider the difference in radii between the outer and
inner tubes in a 1D vdW heterostructure to be the same as the
interlayer distance $d$ of a corresponding 2D vdW heterostructure,
which in the case of 2D MoS$_2$/WS$_2$ heterostructure has a value $d
\approx 6.2$~\AA~\cite{Komsa2013}. For a zigzag nanotube with a chiral
index $(n,0)$, the tube diameter $D$ is related to the chiral index
$n$ as $D = na_0(1+\varepsilon)/\pi$, where $a_0 = |\mathbf{a}_1|$ is
the in-plane lattice constant of the corresponding TMDC monolayer (see
Figure~\ref{fig:fig3}a), and $\varepsilon$ is the circumferential
tensile strain discussed earlier. Based on $D_{\text{outer}} -
D_{\text{inner}} \approx 2d$, we determine that the difference in
chiral index between the outer and inner nanotubes of a 1D vdW
heterostructure is $n_{\text{outer}} - n_{\text{inner}} \approx
12$. Hence, we investigate the band alignment in 1D vdW
heterostructures formed between $(n_{\text{outer}},0)$ and
$(n_{\text{outer}}-12,0)$ nanotubes of MoS$_2$ and WS$_2$.

Figure~\ref{fig:fig6}a shows the alignment between the VBM and CBM of
larger-diameter MoS$_2$$(n_{\text{outer}},0)$ nanotubes and
smaller-diameter WS$_2$$(n_{\text{outer}}-12,0)$ nanotubes as a
function of $n_{\text{outer}}$ and $D_{\text{outer}}$, before coaxial
1D vdW heterostructures are formed. Between such individually
separated nanotubes of MoS$_2$ and WS$_2$, at large tube diameter
$D_{\text{outer}}$, the band alignment belongs to the Type~\Rmnum{2}
category, which is in consistent with MoS$_2$/WS$_2$ 2D vdW
heterostructures formed by stacking of monolayers. As the diameters of
the nanotubes (determined by $n_{\text{outer}}$) decrease, because the
CBM energy level of the smaller-diameter
WS$_2$$(n_{\text{outer}}-12,0)$ nanotube decreases faster with
$n_{\text{outer}}$ than that of the larger-diameter
MoS$_2$$(n_{\text{outer}},0)$ nanotube, the band alignment eventually
becomes Type~\Rmnum{1}, albeit at a small tube diameter of
$D_{\text{outer}}\approx 28$~\AA.

However, when the WS$_2$ and MoS$_2$ nanotubes form coaxial 1D vdW
heterostructures, the positive flexovoltage $\Delta V_{\text{flexo}}$
of the outer MoS$_2$ nanotube, generated in the region where the
smaller-diameter WS$_2$ nanotube is nested, will cause an additional
lowering of the energy levels of the inner tube, thus accelerating the
transition of band alignment from Type~\Rmnum{2} to
Type~\Rmnum{1}. This can be clearly seen from Figure~\ref{fig:fig6}b,
where we have added the additional electrostatic potential energy
shifts $-|e|\Delta V_{\text{flexo}}$, caused by the flexovoltage of
the outer MoS$_2$ nanotube, to the VBM/CBM levels of the inner WS$_2$
nanotube. The band alignment between the coaxial
nanotubes is then replotted. It can be seen that the combined effects of
diameter-dependent band-edge levels and flexovoltage-induced
electrostatic potential shift lead to a transition from
Type~\Rmnum{2} to Type~\Rmnum{1} band alignment in the WS$_2$@MoS$_2$
1D vdW heterostructure at a larger critical diameter of
$D_{\text{outer}} \approx 35$~\AA. The critical transition point
corresponds to a WS$_2$(22,0)@MoS$_2$(34,0) heterostructure.

To confirm the predicted existence of transition in band alignment
from Type~\Rmnum{2} to Type~\Rmnum{1} in WS$_2$@MoS$_2$ 1D vdW
heterostructures, we have explicitly calculated the electronic band
structure of a WS$_2$(10,0)@MoS$_2$(22,0) coaxial heterostructure.
The almost identical lattice constants of MoS$_2$ and WS$_2$ nanotubes
along the tube-axis direction ($T = \sqrt{3}a_0$, with
$a_0\approx3.18$~\AA), as well as the relatively small values of
$n_{\text{outer}}$ and $D_{\text{outer}}$ in this model system, allow
us to build an explicit atomistic representation of the 1D vdW
heterostructure and carry out DFT calculations of its electronic
structure. Note that such calculations would be extremely challenging
if a 1D vdW heterostructure is built from TMDC nanotubes with
incommensurate lattice constants along the tube-axis direction (which
is the case for heterostructures formed between all other Mo- or
W-dichalcogenide nanotubes), as a much larger supercell would be
needed to model a heterostructure in which both the inner and outer
nanotubes are in strain-free state along the translational symmetric
direction. The challenges of direct first-principles calculation of
the electronic structures of 1D vdW heterostructures represent a key
reason that, after using the present small-size model system to
demonstrate the reliability of our approach, we will determine the
band alignment in all other 1D vdW heterostructures using the same
approach, that is by combining the diameter-dependent band energy
evolution of individual TMDC nanotubes with the inter-tube coupling
effect coming from flexovoltage.

The calculated electronic band structure of the
WS$_2$(10,0)@MoS$_2$(22,0) heterostructure is shown in
Figure~\ref{fig:fig6}c. Projection of the electronic states near the
fundamental gap onto the inner and outer MoS$_2$ nanotubes
demonstrates that the predominant contributions to both the VBM and
CBM of the heterostructural system stem from the inner WS$_2$(10,0)
nanotube. This is also explicitly shown in Figure~\ref{fig:fig6}d,
where it can be seen that both the VBM and CBM wavefunctions are
localized on the inner tube. The results thus confirm that the
WS$_2$(10,0)@MoS$_2$(22,0) heterostructure has a Type~\Rmnum{1} band
alignment, in consistent with the prediction in
Figure~\ref{fig:fig6}b.

So far, we have neglected the electronic coupling between the inner
and outer nanotubes in a 1D vdW heterostructure beyond the
flexovoltage effect. We show below that this approximation is
sufficient to determine the band alignment in 1D vdW heterostructures
under most circumstances. With respect to the alignment of CBM, since
the CBM of TMDC nanotubes are all predominantly derived from the $d$
orbitals of the transition-metal (Mo or W) atoms, with only a minor
contribution from the $p_x/p_y$ orbitals of chalcogen (S, Se, or Te)
atoms, the CBM wavefunctions of TMDC nanotubes are well confined
within respective nanotubes. Thus, the electronic wavefunctions of CBM
in inner and outer nanotubes have little overlap, resulting in
negligible effect of hybridization on the CBM energy levels. This
rather weak hybridization of CBM in TMDC heterostructures has been
shown in the case of 2D vdW heterostructures of TMDC
monolayers~\cite{Komsa2013}, and it is also clearly shown for the
WS$_2$@MoS$_2$ 1D vdW heterostructure in
Figure~\ref{fig:fig6}c. Therefore, the CBM band alignment determined
using our approach is rather robust in practice, which is corroborated
by comparing the calculated CBM energy level of the WS$_2$(10,0)
nanotube in the 1D vdW heterostructure in Figure~\ref{fig:fig6}b and
Figure~\ref{fig:fig6}c, where we find that the difference in CBM
energy between our approach and explicit DFT calculation is as small
as 0.035~eV.

\begin{figure*}[t!]
	\centering
	\includegraphics[width=0.82\textwidth]{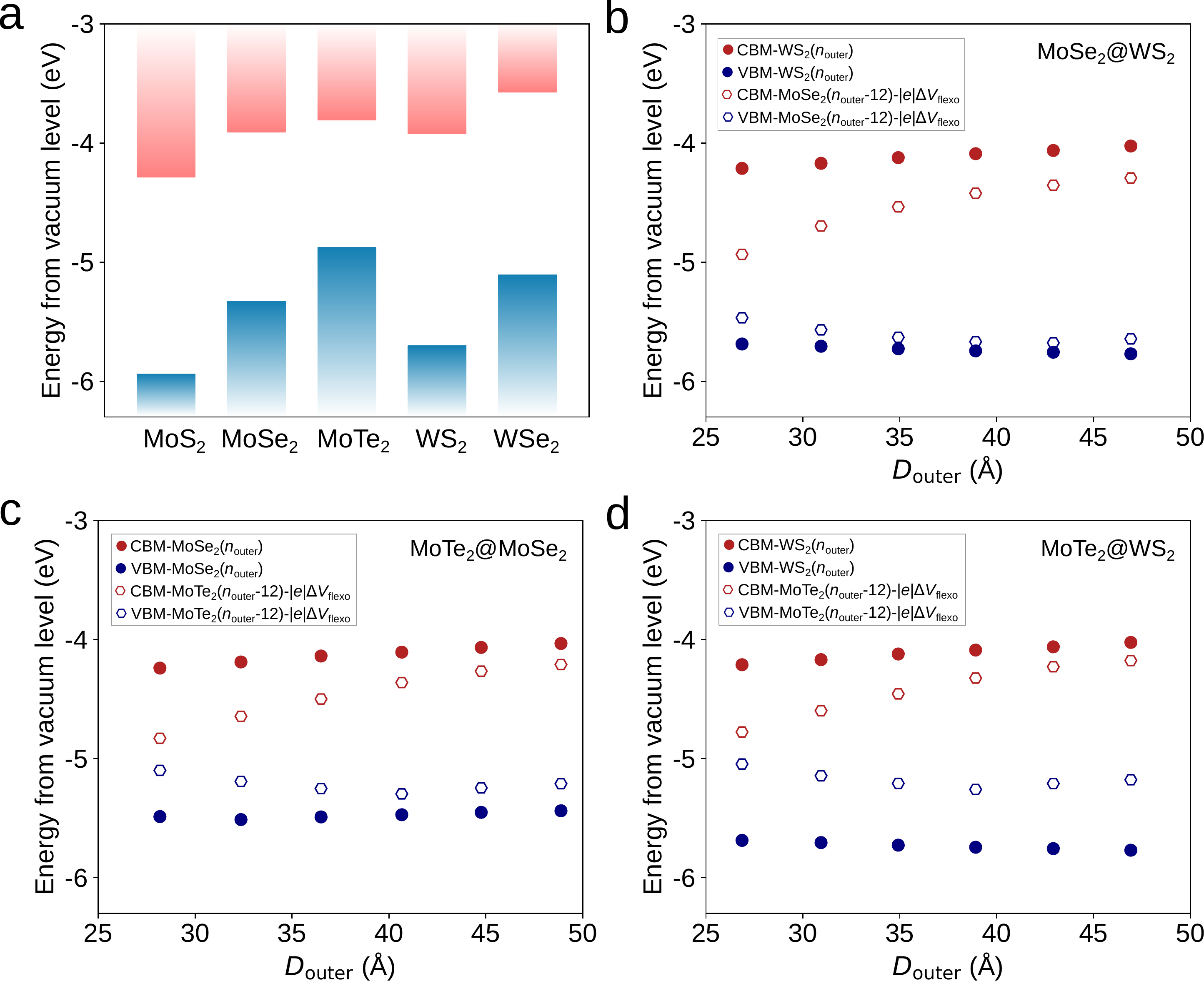}
	\caption{1D vdW heterostructures that exhibit Type~\Rmnum{1}
          band alignment at large tube diameters. (\textbf{a})
          Calculated band alignment between semiconducting Mo- and
          W-dichalcogenide monolayers. Blue and red represent valence
          and conduction band, respectively. (\textbf{b}) Band
          alignment between inner MoSe$_2$ and outer WS$_2$ nanotubes
          in MoSe$_2$@WS$_2$ 1D vdW heterostructures, constructed from
          calculated VBM and CBM energy levels of individual
          WS$_2$$(n_{\text{outer}},0)$ and
          MoSe$_2$$(n_{\text{outer}}-12,0)$ nanotubes, and accounting
          for the additional effect of the flexoelectric potential
          $\Delta V_{\textrm{flexo}}$ generated by the outer WS$_2$
          nanotube on the inner MoSe$_2$
          nanotube. (\textbf{c},\textbf{d}) Similar to (\textbf{b}),
          but for MoTe$_2$@MoSe$_2$ and MoTe$_2$@WS$_2$ 1D vdW
          heterostructures, respectively.}
\label{fig:fig7}
\end{figure*}

The effect of intertube hybridization on the VBM alignment of inner
and outer nanotubes in 1D vdW heterostructures is more subtle than
that of CBM. In a 2D vdW heterostructure of TMDC monolayers,
ref.~\citep{Komsa2013} shows that the hybridization between the
monolayer VBM states at the $K$ point of 2D Brillouin zone, which
mainly derive from the $d_{x^2-y^2} + d_{xy}$ orbitals of
transition-metal atoms, is also negligible. On the other hand, the
hybridization between the VBM states at $\Gamma$, which have
out-of-plane characters and mainly derive from the transition-metal
$d_{z^2}$ and chalcogen $p_z$ orbitals, depends on the relative energy
levels of the VBM states prior to the formation of the 2D vdW
heterostructure. If the VBM at $\Gamma$ of the two monolayers are
close in energy, then hybridization effect can be strong, even pushing
the energy of interlayer ``antibonding'' state above that of the
highest VBM at $K$, as in the case of MoS$_2$/WS$_2$ 2D vdW
heterostructure~\cite{Komsa2013}. However, in our case of 1D vdW
heterostructures of TMDC nanotubes, the strong internal flexoelectric
electric field pointing from the inner side to the outer side of a
nanotube, whose magnitude is estimated to be on the order of $\Delta
V_{\text{flexo}}$ divided by the thickness of the tube wall (up to 1
V/nm in small-diameter nanotubes), will suppress the delocalization of
VBM wavefunctions between inner and outer nanotube. Such electric
field induced localization of electronic states has been shown for
TMDC bilayers under a vertical electrical field, as well as in
double-walled nanotubes~\cite{Ramasubramaniam2011, Wang2021}. Hence,
the effect of hybridization on the energy levels of VBM in 1D vdW
heterostructure is limited. This is again corroborated by the
explicitly calculated band-energy levels of the
WS$_2$(10,0)@MoS$_2$(22,0) heterostructure in Figure~\ref{fig:fig6}c,
where by comparison with Figure~\ref{fig:fig6}b (obtained by
considering only the flexovoltage effect), we determine that the
VBM energy-level difference between the two approaches is also small
(within 0.15~eV).

The computationally proven existence of band-alignment transition from
Type~\Rmnum{2} to Type~\Rmnum{1} in WS$_2$@MoS$_2$ 1D vdW
heterostructures motivates us to investigate all 20 possible types of
1D vdW heterostructures formed between MoS$_2$, MoSe$_2$, MoTe$_2$,
WS$_2$, and WSe$_2$ nanotubes. In Figure~S8 to
Figure~S17, we present the calculated band alignment between
$(n_{\text{outer}},0)$ and $(n_{\text{outer}}-12,0)$ nanotubes of
these TMDCs before and after forming 1D vdW heterostructures.  The
most important finding from these calculations, as shown in
Figure~\ref{fig:fig7}, is that three types of 1D vdW heterostructures,
namely MoSe$_2$@WS$_2$, MoTe$_2$@MoSe$_2$, and MoTe$_2$@WS$_2$,
exhibit Type~\Rmnum{1} band alignment at rather large heterostructural
tube diameters ($D_{\text{outer}}$ around or above 50~\AA), which are
already experimentally accessible~\cite{Xiang2020}. The existence of
Type~\Rmnum{1} band alignment in large-diameter systems are important
not only because such systems are easier to fabricate experimentally,
but also because a large-diameter inner nanotube in the
heterostructure would not have undergone a change in the VBM
character, therefore still exhibiting strong
photoluminescence. Because these three heterostructural systems have
Type~\Rmnum{1} band alignment, electron-hole pairs generated by light
illumination on the outer nanotube would be transferred to the inner
nanotube region and radiatively recombine and emit photons there,
realizing spatially separated photon absorption and emission. As such,
these three types of 1D vdW heterostructure (MoSe$_2$@WS$_2$,
MoTe$_2$@MoSe$_2$, and MoTe$_2$@WS$_2$) could potentially find
important applications in nanoscale optoelectronics.

The reason that these three 1D vdW heterostructural systems can
undergo Type~\Rmnum{2} to Type~\Rmnum{1} band alignment transition at
large tube diameters can be understood by considering the band
alignment of their corresponding 2D monolayers, which are shown in
Figure~\ref{fig:fig7}a. Between any two of MoSe$_2$, MoTe$_2$, and
WS$_2$ monolayers, the difference in their CBM energy levels is small
-- less than than 0.2~eV, while the difference in their VBM energy
levels is relatively large (on the order of 0.5~eV or higher). The
small difference in the CBM level of the three TMDCs has been found to
be quite robust with respect to the computational methodologies
employed~\cite{Liang2013}. Although the three TMDCs form
Type~\Rmnum{2} band alignment in 2D vdW heterostructures, in 1D vdW
heterostructures, if the inner tube belongs to a TMDC that has higher
CBM level than the outer TMDC in the monolayer form, due to the
combined effect of the inner tube experiencing more diameter-induced
lowering of the CBM, as well as the additional lowering of band
energies due to the flexovoltage generated by the outer tube, the CBM
level of the inner tube can be pulled down to such a level that it
becomes lower than the CBM of the outer tube. Meanwhile, because the
VBM energy difference between the monolayers of inner and outer TMDCs
is relatively large, the VBM level of the inner tube remains higher
than that of the outer tube. Consequently, a transition from
Type~\Rmnum{2} to Type~\Rmnum{1} band alignment is readily induced in
these three 1D vdW heterostructural systems.

The flexovoltages $\Delta V_{\text{flexo}}$ in TMDC nanotubes, which
are on the order of 0.15--0.2~eV when the tube diameter is around
50~\AA\ (Figure~S3), is by no means a small number. Even when the tube
diameter increases to 100~\AA, $\Delta V_{\text{flexo}}$ is still on
the order of 0.1~eV. Such large values of flexovoltages alone are
enough to shift the relative order of CBM levels in certain 1D vdW
heterostructures with close CBM energies between individual nanotubes.

Hitherto, we have established a complete framework of band alignment
in 1D vdW heterostructure, by not only fully understanding the
diameter-dependent evolution of the band-edge levels in individual
nanotubes, but also unraveling the crucial effect of flexovoltage in
inducing the transition from Type~\Rmnum{2} to Type~\Rmnum{1} band
alignment in certain 1D vdW heterostructures. Although these results
are built on a solid foundation, we would like to comment on a key
technical aspect of our study. A question that a prospective reader
may ask is whether DFT can accurately predict the band
alignment of these 1D vdW heterostructures. Our answer to this question
is solidly positive. First of all, a critical influencing factor of band
alignment in 1D vdW heterostructures, that is the flexoelectricity
 and flexovoltage effect, is the ground-state property of a TMDC
nanotube, for which DFT has excellent predictive accuracy. Second, although
DFT is known to underestimate the bandgap of semiconductors, a
previous study of band-edge levels in TMDC monolayers using more
sophisticated $GW$ calculations indicates that the
underestimates of bandgap in DFT are relatively uniform across
different TMDCs~\cite{Liang2013}. In particular, the relative
energy differences of CBM levels among different TMDC monolayers
are changed very little when switching the calculation method from DFT
to $GW$. Thus, the results of this work are expected to be robust even
if more sophisticated (but computationally extremely demanding)
methodologies are adopted.

\section*{Conclusions}
In conclusion, through comprehensive first-principles calculations and
theoretical analysis, we have established a complete framework for
understanding the band alignment in 1D vdW heterostructures of coaxial
TMDC nanotubes. We have shown that the CBM levels of individual TMDC
nanotubes exhibit a rapid and monotonic lowering when the tube
diameter is reduced below 50--60~\AA, whereas the VBM of TMDC nanotubes
display an initial lowering before rising, with the transition
diameter varies from $\sim$50~\AA\ in MoS$_2$ and WS$_2$ nanotubes to
$\sim$30~\AA\ in MoSe$_2$, WSe$_2$, and MoTe$_2$ nanotubes. These
properties can be fully explained in terms of curvature-induced
flexoelectricity and the associated electrostatic potential effect, as
well as the intrinsic circumferential tensile strain and bending strain
within the TMDC nanotubes. Quantum confinement effect is found to play
a negligible role on the band-edge evolution.

When the diameter of a TMDC nanotube is above 50~\AA, the band-edge
level evolution of both VBM and CBM in the nanotube is predominantly
determined by the flexoelectricity-induced electrostatic potential effect, which
lowers both the CBM and VBM levels of TMDC nanotubes by 50--100~meV in
the diameter range of 50--100~\AA. When the tube diameter is below
50~\AA, the effect of circumferential tensile strain, which can reach
$\sim$2\% in a TMDC nanotube with a diameter of $\sim$30~\AA, is the
main source responsible for the rapid lowering of the CBM and the
non-monotonic change of the VBM. We have shown that, the transition of
the VBM of TMDC nanotube from downward to upward shifting as
diameter reduces is caused by a change of the VBM orbital character
from predominantly in-plane transition-metal $d_{x^2-y^2} + d_{xy}$
orbitals to out-of-plane transition-metal $d_{z^2}$ plus the chalcogen
$p_z$ orbitals. The transition in the VBM orbital
character of small-diameter TMDC nanotubes can be readily understood
using the scheme of Brillouin-zone folding from the 2D band
structures of uniaxially strained monolayer to the 1D band structures
of nanotubes. We also show that the transition in the orbital
character of VBM leads to direct-to-indirect bandgap transition in
small-diameter armchair or chiral nanotubes, as well as
photoluminescence quenching in zigzag nanotubes.

Building on the comprehensive understanding of the diameter-dependent
evolution of the band-edge levels in individual TMDC nanotubes, we
have investigated the band alignment in all 20 possible types of 1D
vdW heterostructures formed between Mo- and W-dichalcogenide
nanotubes. We show that the large flexoelectric voltages generated by
outer nanotubes on the inner nested nanotubes play a crucial role in
determining the band alignment of TMDC nanotubes in 1D vdW
heterostructures. The combination of diameter-dependent band-edge
levels, as well as the flexovoltage effect, leads to a transition from
Type~\Rmnum{2} to Type~\Rmnum{1} band alignment in multiple 1D vdW
heterostructural systems. In particular, we identify three 1D vdW
heterostructural systems, namely MoSe$_2$@WS$_2$, MoTe$_2$@MoSe$_2$,
and MoTe$_2$@WS$_2$, that already exhibit Type~\Rmnum{2} to
Type~\Rmnum{1} transition when the outer-tube diameter is around or
above 50~\AA. These large-diameter 1D vdW heterostructural systems
should still exhibit photoluminescence, and it is expected that they
could be fabricated experimentally with relative ease, making these 1D
heterostructural systems attractive for nanoscale optoelectronic
applications.

Altogether, our work lays down a key foundation for understanding
band alignment in 1D vdW heterostructures and paves the way for
rational design of TMDC-based 1D vdW heterostructures.

\section*{Method}
First-principles calculations of the structural and electronic
properties of TMDC nanotubes, monolayers, and 1D vdW heterostructures
are carried out using DFT within the generalized gradient
approximation (GGA) and Perdew-Burke-Enzerhof (PBE)
exchange-correlation functional~\cite{Perdew1996}, as implemented in
the Vienna Ab initio Simulation Package (VASP)~\cite{Kresse1996}. The
electron-ion interaction is described using the projector
augmented-wave method~\cite{Blochl1994, Kresse1999}. Electron
wavefunctions are expanded in a plane-wave basis set with a consistent
cut-off energy of 350~eV, which is sufficient to converge total energy
calculations within 3~meV per atom. Atomistic models of TMDC nanotubes
are built from TMDC monolayers with fully relaxed lattice
constants. To avoid the spurious interaction between periodic image
cells in plane-wave DFT calculations, a vacuum space of at least
15~\AA\ perpendicular to the tube-axis direction is added in the
supercell. The atomic positions of nanotubes are fully relaxed until
the maximum force on each atom is less than 0.01~eV/\AA.  For TMDC
monolayers, the 2D Brillouin zones are sampled using a $12 \times 12
\times 1$ Monkhorst-Pack mesh, while the 1D Brillouin zones of TMDC
nanotubes and heterostructures are sampled using a $1 \times 5 \times
1$ $\mathbf{k}$-point mesh, with the non-unity dimension corresponding
to the axial direction of nanotubes or heterostructures. Spin-orbit
coupling (SOC) is not included as spin splitting is symmetry forbidden
in both zigzag and armchair (non-chiral)
nanotubes~\cite{Milivojevic2020}. We have also compared the
electronic band structures of a MoS$_2$(14,0) nanotube calculated with
and without SOC, as shown in Figure~S18. The result confirms
the absence of spin-splitting in the zigzag nanotube and that the
energy difference between the band-edge levels calculated with and
without SOC is negligible.

The vacuum energy level corresponding to each nanotube, monolayer, or
1D vdW heterostructural system is obtained by calculating the
converged Hartree potential energy in the vacuum region far away from
the modeled subject.  For band alignment study, the energies of all
band-edge states are aligned with respect to vacuum level.

\section*{Acknowledgements}
We gratefully acknowledge the support by NSFC under Project
No. 62004172. The work of W.L. is partially supported by Research
Center for Industries of the Future at Westlake University under Award
No. WU2022C041. The authors thank Drs. C.-M. Dai, J.-Q. Wang, and C. Hu
for helpful discussions and the HPC Center of Westlake
University for technical assistance.

\bibliography{ref}

\end{document}